\documentclass[modern]{aastex631}
\newcommand{\change}[1]{#1}

\usepackage{amsmath}
\usepackage{amssymb}
\usepackage{bbold}

\begin{document}

\title{Bayesian model reconstruction based on spectral line observations}

\author[0000-0001-5887-8498]{Frederik De Ceuster}
\affiliation{Institute of Astronomy, Department of Physics \& Astronomy, KU Leuven, \\
Celestijnenlaan 200D, 3001 Leuven, Belgium}

\author[0000-0002-7808-9039]{Thomas Ceulemans}
\affiliation{Institute of Astronomy, Department of Physics \& Astronomy, KU Leuven, \\
Celestijnenlaan 200D, 3001 Leuven, Belgium}

\author[0000-0002-5342-8612]{Leen Decin}
\affiliation{Institute of Astronomy, Department of Physics \& Astronomy, KU Leuven, \\
Celestijnenlaan 200D, 3001 Leuven, Belgium}

\author[0000-0002-1283-6038]{Taïssa Danilovich}
\affiliation{School of Physics \& Astronomy, Monash University, \\
Clayton, Victoria, Australia}
\affiliation{ARC Centre of Excellence for All Sky Astrophysics in 3 Dimensions (ASTRO 3D), \\
Clayton, Victoria, Australia}
\affiliation{Institute of Astronomy, Department of Physics \& Astronomy, KU Leuven, \\
Celestijnenlaan 200D, 3001 Leuven, Belgium}

\author[0000-0003-1954-8749]{Jeremy Yates}
\affiliation{Department of Computer Science, University College London, \\
WC1E 6EA, London, United Kingdom}

\correspondingauthor{Frederik De Ceuster}
\email{frederik.deceuster@kuleuven.be}




\begin{abstract}
Spectral line observations encode a wealth of information.
A key challenge, therefore, lies in the interpretation of these observations in terms of models to derive the physical and chemical properties of the astronomical environments from which they arise.
In this paper, we present \textsc{pomme}: an open-source \textsc{Python} package that allows users to retrieve 1D or 3D models of physical properties, such as chemical abundance, velocity, and temperature distributions of (optically thin) astrophysical media, based on spectral line observations.
We discuss how prior knowledge, for instance, in the form of a steady-state hydrodynamics model, can be used to guide the retrieval process, and demonstrate our methods both on synthetic and real observations of cool stellar winds.
\end{abstract}



\section{Introduction} \label{sec:intro}
A typical problem in astronomy is that for most of our observations we are restricted to the plane of the sky.
As a result, these observations are always mere projections containing only partial information about the distribution of the physical properties of the observed object, especially along the line of sight.
Luckily, in some frequency bands, the observed media are optically thin, such that we receive radiation from the entire medium along the line of sight, and thus we can, at least in principle, extract information from it.
This is especially the case for spectral line interactions caused by transitions between the quantized energy levels of atoms and molecules in the medium.
We will focus specifically on rotational transition lines which are excited in many astrophysical environments and can easily be resolved individually.

Spectral lines encode a wealth of information.
Their characteristic frequencies allow us to identify the chemical species that produce them, while their particular shapes allow us to deduce the physical state of the medium.
Their narrow extent in frequency space makes them particularly sensitive to Doppler shifts, such that velocity gradients (along the line of sight) encode information about the distribution of physical properties (along the line of sight) in the spectral energy distribution.
In this paper, we provide the tools to decode this information in an informed way.

We present an open-source \textsc{Python} package, called \textsc{pomme}\footnote{The source code can be found at: \url{https://github.com/Magritte-code/pomme}.}, to create 1D or 3D models of physical properties, such as chemical abundance, velocity, and temperature distributions, based on spectral line observations.
We built these tools specifically to help us with the interpretation of complex observations, such as, the intricate stellar winds around evolved stars \citep[see e.g.][]{mauron_imaging_2006, ramstedt_wonderful_2014, kervella_alma_2016, decin_substellar_2020}.
Nevertheless, they are also readily applicable to other astrophysical objects, such as protoplanetary disks \cite[see e.g.][]{andrews_disk_2018, oberg_molecules_2021}, or supernova remnants \cite[see e.g.][]{milisavljevic_bubble-like_2015, abellan_very_2017}.
In the reconstruction process, we aim to leverage as much as possible any prior knowledge about the astrophysical object.
In particular, we will demonstrate how simple steady-state hydrodynamics can be enforced on the model parameters by introducing an appropriate prior.
Despite the use of prior knowledge, there will always remain some degeneracy in the possible reconstructions of an object.
We aim to capture this degeneracy, or, more generally, the uncertainty associated with the reconstruction, with a probability distribution over the possible reconstructions, and use the mode (i.e.\ maximum) of this distribution as estimate for the reconstruction.

There is a long and intricate history of reconstruction methods.
Many of these methods have been invented and re-invented by different authors in different contexts, ranging from astronomy and geophysics, all the way to medical imaging, and, more generally, applied mathematics.
Depending on the context, these methods are also known by other names such as inversion or deprojection. 
We will only provide some brief highlights of previous work in the context of astronomy.

Already more than a century ago, \cite{von_zeipel_catalogue_1908} used the Abel transform to make spherical deprojections of globular clusters.
Later, this method was extended for cylindrical symmetric distributions and used to create deprojections, for instance, of supernova remnants \citep{kaastra_deprojection_1989} and cometary atmospheres \citep{hubert_inversion_2016}.
Also the solar physics community has a long and rich history of inversion techniques \citep[see e.g.\ the reviews by][and the references therein]{del_toro_iniesta_inversion_2016, dela_cruz_rodriguez_radiative_2017}.
The environments of evolved stars, such as asymptotic giant branch (AGB) stars and red super giants (RSG), provide a particularly interesting use case for reconstruction methods.
While in some cases one can find appropriate analytical or numerical hydrodynamics models that match the observations \citep[see e.g.][]{maercker_unexpectedly_2012, homan_alma_2018, homan_unusual_2018, chen_three-dimensional_2016, lee_formation_2022, danilovich_chemical_2024}, the cases with the most complex morpho-kinematics can probably only be interpreted using reconstruction methods \citep[see e.g.][]{guelin_irc_2018, montarges_noema_2019, coenegrachts_unusual_2023}, since without a systematic approach, one cannot hope to find a model that matches the observations.
Therefore, in this paper, we will illustrate our methods with examples of evolved stars.

A key problem has always been the ill-posedness of the reconstruction problem, i.e.\ the lack of information in observations to fully constrain the reconstructed model.
\cite{rybicki_deprojection_1987} derived hard mathematical constraints on the reconstructive power of deprojection methods.
However, \cite{palmer_deprojection_1994} showed that, with reasonable assumptions on the reconstructed spatial distributions, these constraints can mostly be evaded. In mathematics literature, these assumptions would now be classified as regularisation methods, or, in the Bayesian setting, as the choice of a prior \citep[see e.g.][]{bertero_introduction_2021}.
The probabilistic approach to deal with the lack of information by explicitly modelling the uncertainty in the reconstruction already appears in the seminal paper by \cite{lucy_iterative_1974} and turned out to be very fruitful.
This Bayesian approach was developed further in the solar physics community by \cite{asensio_ramos_bayesian_2007} along with novel regularisation methods, for instance, sparsity constraints on the reconstructed models \citep[see][]{asensio_ramos_sparse_2015}.

In recent years, with the rise of machine learning techniques, several innovative data-driven solution methods for reconstruction problems have appeared, such as amortized neural posterior estimation \citep{asensio_ramos_approximate_2022} and normalizing flows \citep{diaz_baso_bayesian_2022, ksoll_deep_2023}.
These methods aim to learn how to reconstruct particular types of objects based on examples of synthetic observations and their corresponding models.
In contrast, our approach is rather classical, mostly, due to the computational cost of generating the required amount of complete enough example models for the stellar wind applications that we have in mind.
Nevertheless, we will show how our classical method can also help to build better data-driven reconstruction tools.
Furthermore, machine learning techniques have also been used to accelerate the reconstruction process by emulating the most time consuming parts of the necessary radiation transport with efficient neural networks \citep[see e.g.][]{de_mijolla_incorporating_2019, vicente_arevalo_accelerating_2022}.
Although, currently, we do not yet use such techniques in \textsc{pomme}, we built our software with this kind of future improvements in mind.

This paper is organised as follows.
In Section \ref{sec:methods}, we describe the reconstruction problem and our solution method together with its software implementation.
The numerical details of the implementation are diverted to Appendix \ref{app:impl}.
In Section \ref{sec:tests}, we test our methods by reconstructing synthetic observations of known models, and in Section  \ref{sec:applications} we present a first application by reconstructing real observations.
We discuss the current limitations and future work in Section \ref{sec:discussion}, and, finally, we present our conclusions in Section \ref{sec:conclusions}.

\section{Methods} \label{sec:methods}
First, we describe the forward problem of spectral line formation and explain how, within the assumptions of our model, the physical properties of the astrophysical medium in combination with the instrumentation properties of the detectors lead to the spectral line images we observe.
Next, we describe how we can solve the inverse problem of reconstructing a model for the physical properties of the object, given spectral line observations.
Finally, we discuss the implementation of these methods in the open-source \textsc{Python} package that we developed, called \textsc{pomme}.

\subsection{Forward problem: spectral line formation \& observation}
The forward problem describes how the physical properties of an (astrophysical) medium give rise to a spectral line observation.
We collectively represent the physical properties of an object by a model vector, $\boldsymbol{m}$, and we represent an observation by a vector, $\boldsymbol{o}$.
Hence, the forward problem can be thought of as a function, $f$, that maps a model vector, $\boldsymbol{m}$, to the corresponding observation vector,
\begin{equation}
    \boldsymbol{o} \ = \ f\left( \boldsymbol{m} \right) .
\end{equation}
We distinguish two stages in the forward function, such that, $f\left( \boldsymbol{m} \right) \ = \ f_{2}\left( f_{1}\left( \boldsymbol{m} \right) \right)$.
The first part, $f_{1}$, describes the spectral line formation, i.e.\ the radiative processes that determine the amount of electromagnetic radiation that emanates from the medium and travels towards our detectors.
This is described in Section \ref{subsubsec:spectrallineformation}.
The second part, $f_{2}$, describes the observation and includes the instrumentation effects on the observed signal. This is described in Section \ref{subsubsec:observationalinstrumentationeffects}.

\subsubsection{Spectral line formation} \label{subsubsec:spectrallineformation}
Throughout this paper, we use Cartesian coordinates, $\boldsymbol{x}=(x,y,z)$, and assume the line of sight to be along the positive $z$-axis, such that the plane of the sky corresponds to the $(x,y)$-plane.
Furthermore, we will assume that the observer is located at $z=0$, and that the model box has a depth, $L$, along the $z$-axis.
Then, in the absence of radiation scattering, considering a single ray through the model (resulting in a single pixel in the observed image), the observed intensity at $\boldsymbol{x}=(x,y,0)$, at frequency, $\nu$, reads,
\begin{equation}
\begin{split}
I_{\text{obs}}(\nu; x, y)
\ =& \ \,
I_{\text{bdy}}(\nu; x, y) \, e^{-\tau_{\text{obs}}(\nu; \, x, y, L)} \\
& \ + \
\int_{0}^{L} \text{d}z \ \eta\big( \nu_{\text{com}}(\nu; \, x, y, z); \, x, y, z \big) \, e^{-\tau_{\text{obs}}(\nu; \, x, y, z)} ,
\label{eq:formalsolution}
\end{split}
\end{equation}
where we defined the intensity of the incoming radiation at the boundary, $I_{\text{bdy}}(\nu; x, y)$, and the optical depth, $\tau_{\text{obs}}$, in the observer frame between $(x,y,0)$ and $(x,y,z)$, as,
\begin{equation}
\tau_{\text{obs}}(\nu; x, y, z)
\ \equiv \
\int_{0}^{z} \text{d}z' \ \chi\big( \nu_{\text{com}}(\nu; \, x, y, z'); \, x, y, z' \big) .
\label{eq:opticaldepth}
\end{equation}
Since we look along the $z$-axis, the non-relativistic Doppler-shifted frequency in the co-moving frame is given by,
\begin{equation}
    \nu_{\text{com}}(\nu, \boldsymbol{x}) \ = \ \left( 1 + \frac{v_{z}(\boldsymbol{x})}{c} \right) \nu ,
\end{equation}
where, $v_{z}(\boldsymbol{x})$, is the component of the velocity field along the line of sight, which we earlier already chose to be along the $z$-axis, and, $c$, is the speed of light.
Equations (\ref{eq:formalsolution}) and (\ref{eq:opticaldepth}), known as the formal solution to the radiative transfer equation, express the monochromatic intensity, $I_{\text{obs}}(\nu; x, y)$, observed along the $z$-axis, at $z=0$, in terms of the monochromatic emissivity, $\eta(\nu; \boldsymbol{x})$, and opacity, $\chi(\nu; \boldsymbol{x})$, throughout the medium, along the line of sight.
The numerical solution of these integrals is discussed in Appendix \ref{app:impl}.
When evaluating the emissivity and opacity, to account for Doppler shifts caused by the macroscopic\footnote{This excludes the microscopic thermal and turbulent motions which we model by the line profile.} motion of the medium along the line of sight, we shift all frequencies to the co-moving frame.

Considering only a single spectral line transition between the quantized energy levels denoted by $i$ and $j$, the corresponding local monochromatic line emissivity, $\eta_{ij}(\nu; \boldsymbol{x})$, and opacity, $\chi_{ij}(\nu; \boldsymbol{x})$, can be written as,
\begin{align}
\eta_{ij}(\nu; \boldsymbol{x}) \ &= \  \eta_{ij}(\boldsymbol{x}) \, n(\boldsymbol{x}) \, \phi_{ij}(\nu; \boldsymbol{x}) , \label{eq:emi} \\
\chi_{ij}(\nu; \boldsymbol{x}) \ &= \  \chi_{ij}(\boldsymbol{x}) \, n(\boldsymbol{x}) \, \phi_{ij}(\nu; \boldsymbol{x}) . \label{eq:opa}
\end{align}
Here, $n(\boldsymbol{x})$, denotes the number density of the chemical species producing the line, and $\phi_{ij}(\nu; \boldsymbol{x})$ is the line profile function  describing the spread of the spectral line in frequency space.
In this paper, we will assume a Gaussian line profile function predominantly caused by Doppler shifts due to the thermal and turbulent motions (along the line of sight) in the medium,
\begin{equation}
    \phi(\nu; \boldsymbol{x})
    \ = \
    \frac{1}{\delta\nu_{ij}(\boldsymbol{x}) \sqrt{\pi}} \ \exp \left( - \left(\frac{\nu - \nu_{ij}}{\delta \nu_{ij}(\boldsymbol{x})}\right)^{2} \right) ,
\label{eq:profile}
\end{equation}
in which the local line width, $\delta\nu_{ij}(\boldsymbol{x})$, is defined as,
\begin{equation}
    \delta\nu_{ij}(\boldsymbol{x})
    \ = \
    \frac{\nu_{ij}}{c} \sqrt{\frac{2 k_{\text{B}}T(\boldsymbol{x})}{m_{\text{spec}}} \ + \ v_{\text{turb}}^{2}(\boldsymbol{x})} .
\end{equation}
The Gaussian line profile is centred around the line frequency, $\nu_{ij}$. The line width is determined by a thermal component characterised by the local gas temperature, $T(\boldsymbol{x})$, and the molecular mass, $m_{\text{spec}}$, of the chemical species producing the line, and a turbulent component characterised by a local turbulent velocity, $v_{\text{turb}}(\boldsymbol{x})$.
The two remaining components in equations (\ref{eq:emi}) and (\ref{eq:opa}) are the line emissivity and opacity,
\begin{align}
\eta_{ij}(\boldsymbol{x}) \ &= \  \frac{h \nu_{ij}}{4 \pi} \, p_{i}(\boldsymbol{x}) A_{ij}, \\
\chi_{ij}(\boldsymbol{x}) \ &= \  \frac{h \nu_{ij}}{4 \pi} \, \Big( p_{j}(\boldsymbol{x}) B_{ji} \ - \ p_{i}(\boldsymbol{x}) B_{ij} \Big) ,
\end{align}
which are determined by the Einstein $A_{ij}$, $B_{ji}$, and $B_{ij}$ coefficients quantifying the rates of spontaneous emission, absorption, and stimulated emission respectively.
Note that stimulated emission is modelled as negative absorption.
The $p_{i}(\boldsymbol{x})$ denote the relative populations of the energy levels and represent the local quantum mechanical state of the medium.
These are normalised such that $\sum_{i} p_{i}(\boldsymbol{x}) = 1$, in which the sum iterates over all energy levels of the chemical species under consideration.

Assuming that the medium is in local thermodynamic equilibrium (LTE), the local level populations are completely determined by the local gas temperature, $T(\boldsymbol{x})$,
\begin{equation}
    p_{i}(\boldsymbol{x})
    \ = \
    \frac{g_{i}}{Z(\boldsymbol{x})} \,
    \exp \left(-\frac{E_{i}}{k_{\text{B}} T(\boldsymbol{x})} \right) ,
\end{equation}
where $g_{i}$ and $E_{i}$ denote the statistical weight and the energy of level, $i$, respectively, and the normalisation factor, $Z(\boldsymbol{x})$, is defined such that the local level populations are normalised ($\sum_{i} p_{i}(\boldsymbol{x}) = 1$).
Hence, assuming LTE, the spectral line model is determined completely by the local gas temperature, $T(\boldsymbol{x})$, the number density, $n(\boldsymbol{x})$, the turbulent velocity, $v_{\text{turb}}(\boldsymbol{x})$, and the $z$-component of the velocity, $v_{z}(\boldsymbol{x})$.
All other parameters, such as the radiative constants, can be found in data bases.
In this paper, we use the Leiden Atomic and Molecular Database\footnote{The database can be found at: \url{https://home.strw.leidenuniv.nl/~moldata/.}} \citep[LAMDA;][]{schoier_atomic_2005}.

It should be emphasised that, although all these parameters are encoded in the spectral lines, this does not mean that they also can be retrieved unambiguously.
Note that the information of all distributions along the line of sight is encoded in the frequency-dependence of the intensity in a single pixel, and thus, without further assumptions, some information will necessarily be lost.

\subsubsection{Instrumentation effect}
\label{subsubsec:observationalinstrumentationeffects}
In the previous section, we presented a theoretical model for spectral line formation.
In practice, however, spectral line observations are affected by instrumentation effects, such as binning and noise, the particular form of which highly depends on the way the object is observed.

High spatial and spectral resolution observations are typically obtained using interferometry, for which the situation is even more complicated, since there are spatial scale-dependent effects which cause the smallest and largest structures to be unresolved in the resulting images.
These effects must carefully be taken into account, as they (in part) determine which information about the spatial distribution of the physical properties is encoded in the observations and which is not.
At this point, we should clarify what we mean with interferometric data and distinguish between two types.
The first type are the visibilities, i.e.\ the correlations between the coherent electromagnetic waves, measured in several frequency bins for the different pairs of antennas in the array \citep[see e.g.][]{thompson_fundamentals_1999}.
These visibilities are often further processed into images that map the actual brightness in the plane of the sky for each frequency bin.
This is what we distinguish as the second type of interferometric data.
The conversion of the sky brightness maps from the visibilities can be achieved, for instance, with the \textsc{clean} algorithm \citep[][]{hogbom_aperture_1974}, as implemented, for instance, in \textsc{CASA} \citep{the_casa_team_casa_2022}.
This conversion is in itself already an inverse problem.
Therefore, when reconstructing a model from interferometric observations, it makes sense to start directly from the visibilities.
However, since visibilities are more difficult to interpret than sky brightness maps, we consider both options.

The conversion of the observed intensity in the plane of the sky, $I_{\text{obs}}(\nu; x, y)$, to visibilities, $V(\nu;u,v)$, can be expressed as \citep{thompson_fundamentals_1999},
\begin{equation}
    V(\nu; u, v) 
    \ = \
    \frac{1}{d^{2}}
    \int_{-\infty}^{+\infty} \text{d}x
    \int_{-\infty}^{+\infty} \text{d}y \
    \hat{A}(x,y) \, I_{\text{obs}}(\nu; x, y) \exp\big( -2\pi i \left( u x + v y \right) \big)
\label{eq:visibilities}
\end{equation}
in which $\hat{A}(x,y) \equiv A(x,y) / A_{0}$, is the antenna response function, normalised by, $A_{0}$, the response at the centre of the beam, $d$, is the distance between the object and the observer, and $u$ and $v$ are the components of the projected distance vectors, often referred to as baselines, between pairs of antennas in the array.
Since there are only a finite number of antennas and thus a finite number of antenna pairs, the visibilities are only known at a finite number of baseline samples, $\{(u_{d},v_{d})\}_{d=1}^{N_{d}}$.
This sampling, which depends on the particular antenna configuration, causes a loss of information about the spatial distribution of the source in an intricate way.
From equation (\ref{eq:visibilities}), we can see that the transformation of the intensity in the plane of the sky into visibilities is essentially a 2D Fourier transform.
Although the Fourier transform is invertible, the fact that we can only sample the visibilities for a finite number of baselines, makes that the combined transformation from a sky brightness map to a finite set of visibilities is not invertible.

In principle, we could use the exact pipeline to model the instrumentation effects in our forward model, for instance, using \textsc{CASA} \citep{the_casa_team_casa_2022}.
However, later, in our reconstruction algorithm,
we want the implementation of the forward function, $f$, to be automatically differentiable using the \textsc{autograd} functionality in \textsc{PyTorch} \citep[][see also Sect.\ \ref{subsec:implementation}]{paszke_automatic_2017, paszke_pytorch_2019}.
Therefore, we implemented the map (\ref{eq:visibilities}) ourselves in \textsc{PyTorch} using a fast Fourier transform (FFT), following the \textsc{Galario}\footnote{The source code can be found at: \url{https://github.com/mtazzari/galario}.} package for visibility modelling \citep{tazzari_galario_2018}.

Alternatively, one may want to reconstruct 3D physical models from previously obtained sky brightness maps.
In that case, it is important to consider the beam, i.e.\ a kernel that models the spatial spread of the intensity in the plane of the sky, and the antenna response function.
\begin{equation}
    \tilde{I}_{\text{obs}}(\nu; x, y)
    \ = \
    \frac{1}{d^{2}}
    \int_{-\infty}^{+\infty} \text{d}x'
    \int_{-\infty}^{+\infty} \text{d}y' \
    \hat{A}(x,y) \,
    I_{\text{obs}}(\nu; x', y') \,
    B(x-x',y-y') ,
\label{eq:beam}
\end{equation}
in which $B(x, y)$ is the beam kernel, which we assume to be a 2D Gaussian, centred around $(0,0)$, and, $d$, is again the distance between the object and the observer.

It should be emphasised that in both processes of line formation and observation information is lost.
Mathematically, this implies that the function, $f$, is not invertible.
In the next section, we describe how, with a probabilistic approach, we can circumvent this problem and obtain a probabilistic inverse.

\subsection{Inverse problem: probabilistic 3D reconstruction}
\label{subsec:inverse}
The entire forward problem of spectral line formation and observation, as described above, can also be formulated from a probabilistic point of view, as determining the probability distribution, $p(\boldsymbol{o} \, | \, \boldsymbol{m})$, over all possible observations $\boldsymbol{o}$, given a model $\boldsymbol{m}$.
Classical numerical methods typically only consider a single solution, $\boldsymbol{o} = f(\boldsymbol{m})$, which corresponds to a Dirac delta distribution, $p(\boldsymbol{o} \, | \, \boldsymbol{m}) = \delta\big(\boldsymbol{o} - f(\boldsymbol{m})\big)$.
With probabilistic numerical methods, however, the entire probability distribution can be obtained \citep[see e.g.][]{de_ceuster_radiative_2023}.
This probabilistic approach is key, for instance, to quantify uncertainties in the forward problem, but can also help to solve the (classically non-invertible) inverse problem.
Although we might not be able to invert the forward function, $f$, we can always determine the probability of the inverse situation, $p\left(\boldsymbol{m} \, | \, \boldsymbol{o}\right)$, with Bayes' rule \citep[see e.g.][]{asensio_ramos_bayesian_2007, stuart_inverse_2010},
\begin{equation}
    p\left(
    \boldsymbol{m} \, | \, \boldsymbol{o}
    \right)
    \ = \
    \frac{
        p\left(\boldsymbol{o} \, | \, \boldsymbol{m}\right)
        p\left( \boldsymbol{m} \right)
    }
    {
        p\left( \boldsymbol{o} \right)
    } .
\end{equation}
This allows us to determine the probability distribution over possible models, $\boldsymbol{m}$, corresponding to an observation, $\boldsymbol{o}$.
Since the denominator does not depend on the model, $\boldsymbol{m}$, we can treat it as a mere normalisation constant and only concentrate on the numerator.
Here, we find the likelihood, $p\left(\boldsymbol{o} \, | \, \boldsymbol{m}\right)$, which is related to the forward problem, and the prior, $p(\boldsymbol{m})$, which encodes our assumptions about the model, prior to the observation. 
Just like the forward model, $f$, can be viewed as the maximum of the likelihood, $p\left(\boldsymbol{o} \, | \, \boldsymbol{m}\right)$, the inverse, i.e.\ a reconstruction of a model based on an observation, can be viewed as the maximum of the posterior, $p\left(\boldsymbol{m} \, | \, \boldsymbol{o}\right)$.
Determining this maximum is equivalent to minimising the negative logarithm of the posterior,
\begin{equation}
    -\log p\left(
    \boldsymbol{m} \, | \, \boldsymbol{o}
    \right)
    \ = \
    -\log p\left(\boldsymbol{o} \, | \, \boldsymbol{m}\right)
    \ - \
    \log p\left( \boldsymbol{m} \right)
    \ + \
    \log p\left( \boldsymbol{o} \right) .
\label{eq:loglikelihood}
\end{equation}
Since we want to minimise this over different models, $\boldsymbol{m}$, but for a fixed observation, $\boldsymbol{o}$, the last term will be constant and thus can be neglected.
In this optimisation problem, we distinguish three types of objectives or loss functions, i.e. the functions we aim to minimise,
\begin{align}
    p\left(\boldsymbol{m} \, | \, \boldsymbol{o}\right)
    \ &\equiv \
    \exp
    \big(
    -\mathcal{L}_{\text{tot}}\left(\boldsymbol{m}, \boldsymbol{o} \right)
    \big) ,
    \label{eq:probtot}\\
    p\left(\boldsymbol{o} \, | \, \boldsymbol{m}\right)
    \ &\equiv \
    \exp
    \big(
    -\mathcal{L}_{\text{rep}}\left(f(\boldsymbol{m}), \boldsymbol{o} \right)
    \big) ,
    \label{eq:probrep} \\
    p\left(\boldsymbol{m}\right)
    \ &\equiv \
    \exp
    \big(
    -\mathcal{L}_{\text{reg}}\left(\boldsymbol{m} \right)
    \big) ,
    \label{eq:probreg}
\end{align}
in which, $\mathcal{L}_{\text{rep}}$ is the reproduction loss, $\mathcal{L}_{\text{reg}}$ is the regularisation loss, and, $\mathcal{L}_{\text{tot}}$, is the total loss.
Each of these loss functions quantifies the deviation from an objective.
Hence, neglecting the last term, equation (\ref{eq:loglikelihood}) can be written as,
\begin{equation}
    \mathcal{L}_{\text{tot}}\left(  \boldsymbol{m},  \boldsymbol{o} \right)
    \ = \
    \mathcal{L}_{\text{rep}}\left(f(\boldsymbol{m}), \boldsymbol{o} \right)
    \ + \
    \mathcal{L}_{\text{reg}}\left(  \boldsymbol{m} \right) .
\label{eq:loss}
\end{equation}
Equations (\ref{eq:probtot}, \ref{eq:probrep}, \ref{eq:probreg}) allow one to translate a probabilistic problem about probability distributions into an optimisation problem with loss functions, and vice versa.

\subsubsection{Reproduction loss / likelihood}
The reproduction loss, $\mathcal{L}_{\text{rep}}$, is a measure on the space of observations that quantifies how badly a model fits the observation by measuring the discrepancy between a synthetic observation of a model, $f(\boldsymbol{m})$, and the true observation, $\boldsymbol{o}$.
In this paper,  we consider a typical reproduction loss given by the mean squared error, weighted by a covariance matrix, $\boldsymbol{\Sigma}$, such that,
\begin{equation}
\mathcal{L}_{\text{rep}}\big(f(\boldsymbol{m}), \boldsymbol{o} \big)
\ = \
\frac{1}{2} \,
\big(f(\boldsymbol{m}) - \boldsymbol{o}\big)^{\text{T}} \
\boldsymbol{\Sigma}^{-1} \,
\big(f(\boldsymbol{m}) - \boldsymbol{o}\big) .
\end{equation}
With this definition of the reproduction loss, the likelihood (\ref{eq:probrep}), up to a normalisation constant, corresponds to a multivariate Gaussian distribution.
This distribution can be used to represent the uncertainty in the observations, where the square root of the diagonal of the covariance matrix models the uncertainty per pixel or visibility.

For simplicity, in this paper, we omitted the noise on the observations, i.e.\ $\boldsymbol{\Sigma} = \mathbb{1}$.
However, we did find that our reconstruction method performs significantly better by splitting the reproduction loss into a averaged and a relative part,
\begin{equation}
    \mathcal{L}_{\text{rep}}\big(f(\boldsymbol{m}), \boldsymbol{o} \big)
    \ = \
    \mathcal{L}_{\text{rep}}\Big( \big\langle f(\boldsymbol{m}) \big\rangle, \, \left\langle\boldsymbol{o}\right\rangle \Big)
    \ + \
    \mathcal{L}_{\text{rep}}\left( \frac{f(\boldsymbol{m})}{\big\langle f(\boldsymbol{m})\big\rangle}, \, \frac{\boldsymbol{o}}{\left\langle \boldsymbol{o}\right\rangle} \right) ,
\label{eq:relative_loss}
\end{equation}
where the brackets, $\langle\cdot\rangle$, denote an arithmetic mean along an axis of the data, for instance, the frequency. For images, when considering the mean along the frequency axis, this implies that we add the loss for the frequency-averaged intensities in each pixel and the loss for the frequency-normalised intensity in each pixel. 
In this way, all pixels contribute equally, at least in the relative part of the loss, i.e.\ the second term in equation (\ref{eq:relative_loss}).
Without this, the algorithm has difficulty reconstructing dimmer regions in the observations, since their contribution to the loss is overpowered by the contributions of brighter regions.

\subsubsection{Regularisation loss / prior}
\label{subsubsec:prior}
The regularisation loss, $\mathcal{L}_{\text{reg}}$, is a measure on the space of models that quantifies how well a model fits our prior assumptions.
We consider a regularisation loss, or a corresponding prior, $p(\boldsymbol{m})$, that consists of different parts, each encoding a different assumption about our model.
Below, we will present all types of regularisation that we consider in this paper.
However, not every assumption will always be necessary.
Different combinations can be used for different reconstructions.
In the applications in Sections \ref{sec:tests} and \ref{sec:applications}, we will indicate which of the following priors were used.

Often when solving inverse problems, to avoid over-fitting, one assumes a certain degree of regularity or smoothness of the solution.
In this paper, for each model parameter, $q(\boldsymbol{x})$, with a spatial dependence, $\boldsymbol{x}$, we use the integrated squared Euclidean norm of its gradient\footnote{Using Plancherel theorem, this is equivalent to integrating the Fourier modes, $\mathcal{F}[q](\boldsymbol{k})$ of $q$, weighted by the squared Euclidean norm of their wave number $\|\boldsymbol{k}\|^{2}$, thus penalising higher-order modes.},
\begin{equation}
    \mathcal{L}_{\text{reg}}[q] \ = \ \int \text{d} \boldsymbol{x} \ \| \nabla q(\boldsymbol{x})\|^{2} ,
\label{eq:reg_smooth}
\end{equation}
to quantify its deviation from smoothness.
The main reason for this particular choice of smoothness measure is its straightforward numerical implementation.

For some observations it can also be useful to make assumptions about symmetries, and, in particular, about spherical symmetry. 
Therefore, we implemented a loss function that can quantify deviations from spherical symmetry.
Given an origin point in the model, $\boldsymbol{x}_{O}$, this loss quantifies the average variance within a predetermined set of spherical shells around the origin point,
\begin{equation}
    \mathcal{L}_{\text{sph}}[q]
    \ = \
    \int_{0}^{\infty} \text{d} r \
    \mathbb{V}\big[q(\boldsymbol{x}) \, \big| \, \|\boldsymbol{x}-\boldsymbol{x}_{O}\|=r \big] .
\end{equation}

Next, we consider a loss that can encode the physical laws that govern the variables in our model, i.e.\ the distributions of $\rho$, $\boldsymbol{v}$, and $T$.
Not every configuration of $\rho$, $\boldsymbol{v}$, and $T$ is equally likely to occur, since we expect any configuration to be the result of hydrodynamic evolution from some initial conditions.
We assume this hydrodynamic evolution to be governed by the conservation of mass, momentum, and energy, which can be formalised, in Eulerian form, as,
\begin{align}
    \frac{\partial \rho}{\partial t} \ + \ \nabla \cdot \left( \rho \, \boldsymbol{v} \right) \ &= \ 0 ,
    \label{eq:hydro1} \\
    \frac{\partial \boldsymbol{v}}{\partial t} \ + \ \left( \boldsymbol{v} \cdot \nabla \right) \boldsymbol{v} \ + \ \frac{1}{\rho} \, \nabla P \ + \ \nabla \Phi \ &= \ 0 ,
    \label{eq:hydro2} \\
    \frac{\partial E}{\partial t} \ + \ \nabla \cdot \big( \left(E + P\right) \boldsymbol{v} \big) \ + \ \Lambda \ &= \ 0 .
    \label{eq:hydro3} 
\end{align}
Here, we defined the total internal energy,
\begin{equation}
    E \ = \ \frac{1}{2} \, \rho \, \boldsymbol{v}^{2} \ + \ \rho \, u,
\label{eq:energy}
\end{equation}
consisting of a kinetic term and a thermal internal energy, $u$, which can be related to the pressure, $P$, assuming an equation of state,
\begin{equation}
    P \ = \ \left( \gamma - 1 \right) \rho \, u ,
\label{eq:eos}    
\end{equation}
where $\gamma$ is the adiabatic index, which is related to the internal degrees of freedom in the gas.
The pressure, $P$, in turn, can be related to the temperature, $T$, through the ideal gas law,
\begin{equation}
    P \ = \ \frac{k_{\text{B}}}{\mu} \, \rho \, T ,
\label{eq:ideal_gas_law}
\end{equation}
in which $k_{\text{B}}$ is Boltzmann's constant and $\mu$ is the mean molecular weight of the gas.
The remaining components, $\Phi$ and $\Lambda$, describe the gravitational potential and the cooling function respectively. 
Often, there is, for instance, a central object for which we have a mass estimate, $M$, and an estimated location, $\boldsymbol{x}_{\text{grav}}$\footnote{Model parameters, such as $M$ and $\boldsymbol{x}_{\text{grav}}$ can also be considered as free parameters that can be fitted to the observations by minimising the loss (maximising the posterior).}.
As a result, we can, for instance, assume a gravitational potential of the form,
\begin{equation}
    \Phi(\boldsymbol{x}) \ = \ - \frac{GM}{\| \boldsymbol{x} - \boldsymbol{x_{\text{grav}}} \|} .
\end{equation}
Note that this ignores the self-gravitation of the density distribution $\rho$.
The cooling function is often more difficult to estimate, since it depends on many other parameters of the astrophysical object.
Without additional prior knowledge, one can assume, for instance, $\Lambda = 0$.

Equations (\ref{eq:hydro1}, \ref{eq:hydro2}, \ref{eq:hydro3}) provide five component equations that describe the time-evolution of the five components of our model variables, $\rho$, $\boldsymbol{v}$, and $T$.
However, we do not aim to describe the entire time evolution, but rather a snapshot in time, based on an observation.
As a result, our models lack time dependence and we cannot simply enforce the hydrodynamic equations (\ref{eq:hydro1}, \ref{eq:hydro2}, \ref{eq:hydro3}) as prior assumptions.
Instead, we want to know, at any given time, what is the most likely state to find our model in, given that its evolution is governed by the hydrodynamic equations.
One way to do this is to consider time averages of each time-dependent variable, $q(t)$, defined as,
\begin{equation}
    \left\langle q \right \rangle_{T}
    \ \equiv \
    \frac{1}{T} \int_{0}^{T} \text{d} t \ q(t) .
\end{equation}
For any bounded function, i.e.\ there exist $q_{\min}, q_{\max} \in \mathbb{R}$, such that $q_{\min} \leq q(t) \leq q_{\max}$, for sufficiently large time intervals, i.e.\ $T \rightarrow \infty$, one can easily show that the time average of the time derivative of the function vanishes, i.e.\
\begin{equation}
    \left|
        \lim_{T \rightarrow \infty}
        \left\langle \frac{\partial q}{\partial t} \right \rangle_{T}
    \right|
    \ = \
    \left|
        \lim_{T \rightarrow \infty}
        \left(
            \frac{1}{T} \int_{0}^{T} \text{d} t \ \frac{\partial q}{\partial t}
        \right)
    \right|
    \ \leq \
    \lim_{T \rightarrow \infty}
    \left(
    \frac{q_{\max} - q_{\min}}{T}
    \right)
    \ = \ 0 .
\end{equation}
As a result, one can argue that the time-averaged state of a model, which we use as an estimator for the expected state of a model at any time, is a steady-state solution, i.e.\ a solution with vanishing time derivatives, $\partial_{t} \rho = 0$, $\partial_{t} \boldsymbol{v} = 0$, and $\partial_{t} T = 0$.
It is also quite intuitive that it is more likely to observe a system in a state in which its time evolution is slow, since it spends comparatively more time in those states.
Assuming a steady state and using equations (\ref{eq:energy}, \ref{eq:eos}, \ref{eq:ideal_gas_law}), the hydrodynamic equations (\ref{eq:hydro1}, \ref{eq:hydro2}, \ref{eq:hydro3}) can be rewritten in terms of our model variables, $\rho$, $\boldsymbol{v}$, and $T$,
\begin{align}
    \nabla \cdot \left( \rho \, \boldsymbol{v} \right) \ &= \ 0, \\
    \left( \boldsymbol{v} \cdot \nabla \right) \boldsymbol{v} \ + \ \frac{k_{\text{B}}T}{\mu} \, \nabla \big( \log \rho  \, + \, \log T \big) \ + \ \nabla \Phi \ &= \ 0, \\
    \rho \, \boldsymbol{v} \cdot \nabla \left( \frac{1}{2} \, \boldsymbol{v}^{2} \ + \ \frac{\gamma}{\gamma-1} \frac{k_{\text{B}}T}{\mu} \right) \ + \ \Lambda \ &= \ 0 .
\end{align}
Since these equations only contain (spatial derivatives of) our model variables, we can enforce them as an assumption on our model by defining the following loss functions,
\begin{align}
    \mathcal{L}_{\rho}(\boldsymbol{m})
    \ &\equiv \
    \int \text{d}\boldsymbol{x} \ \frac{1}{\rho^{2}} \big( \nabla \cdot \left( \rho \, \boldsymbol{v} \right) \big)^{2} ,
    \label{eq:3D_cont} \\
    \mathcal{L}_{v_{i}}(\boldsymbol{m})
    \ &\equiv \
    \int \text{d}\boldsymbol{x} \ \frac{1}{v_{i}^{2}}  \left( \left( \boldsymbol{v} \cdot \nabla \right) v_{i} \ + \ \frac{k_{\text{B}}T}{\mu} \, \nabla_{i} \big( \log \rho  \, + \, \log T \big) \ + \ \nabla_{i} \Phi \right)^{2} , \label{eq:loss_mom}\\
    \mathcal{L}_{E}(\boldsymbol{m}) \ &\equiv \
    \int \text{d}\boldsymbol{x} \ \frac{1}{E^{2}} \left( \rho \, \boldsymbol{v} \cdot \nabla \left( \frac{1}{2} \, \boldsymbol{v}^{2} \ + \ \frac{\gamma}{\gamma-1} \frac{k_{\text{B}}T}{\mu} \right)
    \ + \ \Lambda \right)^{2} . \label{eq:loss_eng}
\end{align}
Note that equation (\ref{eq:loss_mom}) defines three loss functions, one for each coordinate axis.
We included normalisation factors to ensure that all loss functions have the same unit of inverse time.
These loss functions can be combined to define a regularisation loss,
\begin{equation}
    \mathcal{L}_{\text{reg}}\left(  \boldsymbol{m} \right)
    \ = \
    w_{\rho} \, \mathcal{L}_{\rho} \left(  \boldsymbol{m} \right)
    \ + \
    \boldsymbol{w}_{\boldsymbol{v}} \cdot \boldsymbol{\mathcal{L}}_{\boldsymbol{v}}\left(  \boldsymbol{m} \right)
    \ + \
    w_{E} \, \mathcal{L}_{E}\left(  \boldsymbol{m} \right) ,
\label{eq:loss_reg}
\end{equation}
which, using equation (\ref{eq:probreg}), also defines a prior distribution.
The weights, $w_{\rho}$, $\boldsymbol{w}_{\boldsymbol{v}}$, and $w_{E}$, are hyper-parameters of the prior that determine the width of the distribution around each of the steady-state constraints and can, furthermore, be used to weigh their relative importance.
The numerical values of these weights cannot be determined from first principles, but could, at least in principle, be learned from data.
We elaborate on this in the discussion of future work (see Section \ref{subsec:future}).
We should note that the loss originating from the continuity equation (\ref{eq:3D_cont}), in practice, is by far the most useful, since it does not depend on external parameters, like the gravitational potential ($\Phi$) or the cooling function ($\Lambda$), which are often difficult to describe accurately.
Although we did not find any examples in which the other loss functions (\ref{eq:loss_mom}, \ref{eq:loss_eng}) improved the reconstruction, we included them here for completeness.

Many other types of regularisation loss can be considered.
In principle, any equation involving the model parameters, say, $a(\boldsymbol{m})=b(\boldsymbol{m})$, can be enforced on the model as regularisation or prior by including a loss term proportional to a monotonically increasing function of $|a(\boldsymbol{m})-b(\boldsymbol{m})|$.
This can be used, for instance, for non-LTE line radiative transfer, to enforce the statistical equilibrium equations on the level populations
\citep[see also][]{stepan_novel_2022}.
Since this requires an efficient way to compute the mean intensities in the line, which poses a challenge for our current implementation (see Section \ref{subsec:implementation}), we postpone this to future work and limit ourselves here to LTE line radiative transfer.

\subsection{Implementation in \textsc{pomme}}
\label{subsec:implementation}
The methods described above are implemented in the open-source \textsc{Python} package, called \textsc{pomme}\footnote{The source code can be found at: \url{https://github.com/Magritte-code/pomme}.}.
The package greatly benefits from the functionality to handle astronomical data that is provided by \textsc{astropy} \citep{the_astropy_collaboration_astropy_2013, the_astropy_collaboration_astropy_2018, the_astropy_collaboration_astropy_2022} and heavily relies on the data structures and algorithms provided by \textsc{PyTorch} \citep{paszke_automatic_2017, paszke_pytorch_2019}.
Both the forward model, including spectral line formation and instrumentation effects, as well as the reconstruction algorithm are implemented in the \textsc{PyTorch} framework. This offers several advantages:
\begin{itemize}
    \item The entire code base is hardware agnostic, i.e.\ it can efficiently be executed on many-core central processing units (CPUs), as well as graphics processing units (GPUs). Furthermore, \textsc{PyTorch}, provides the option to leverage various specialized hardware accelerators, such as tensor processing units (TPUs).

    \item Since also the forward model is implemented in \textsc{PyTorch}, the \textsc{autograd} functionality \citep{paszke_automatic_2017} makes it automatically differentiable, such that gradients of the output variables with respect to all input variables are available, and vice versa.
    This allows one to derive uncertainties on the results of the forward model just from a single evaluation of the model, while it also allows us to efficiently solve the inverse problem using gradient descent.

    \item \textsc{PyTorch} offers a wide variety of methods to solve optimisation problems, such as our reconstruction task. Furthermore, having all simulation components in the \textsc{PyTorch} framework makes it easy to employ machine learning methods, for instance, to accelerate parts of the forward model by emulation, as we will explore in future work. 
\end{itemize}

\begin{figure}
    \centering
    \includegraphics[width=0.8\linewidth]{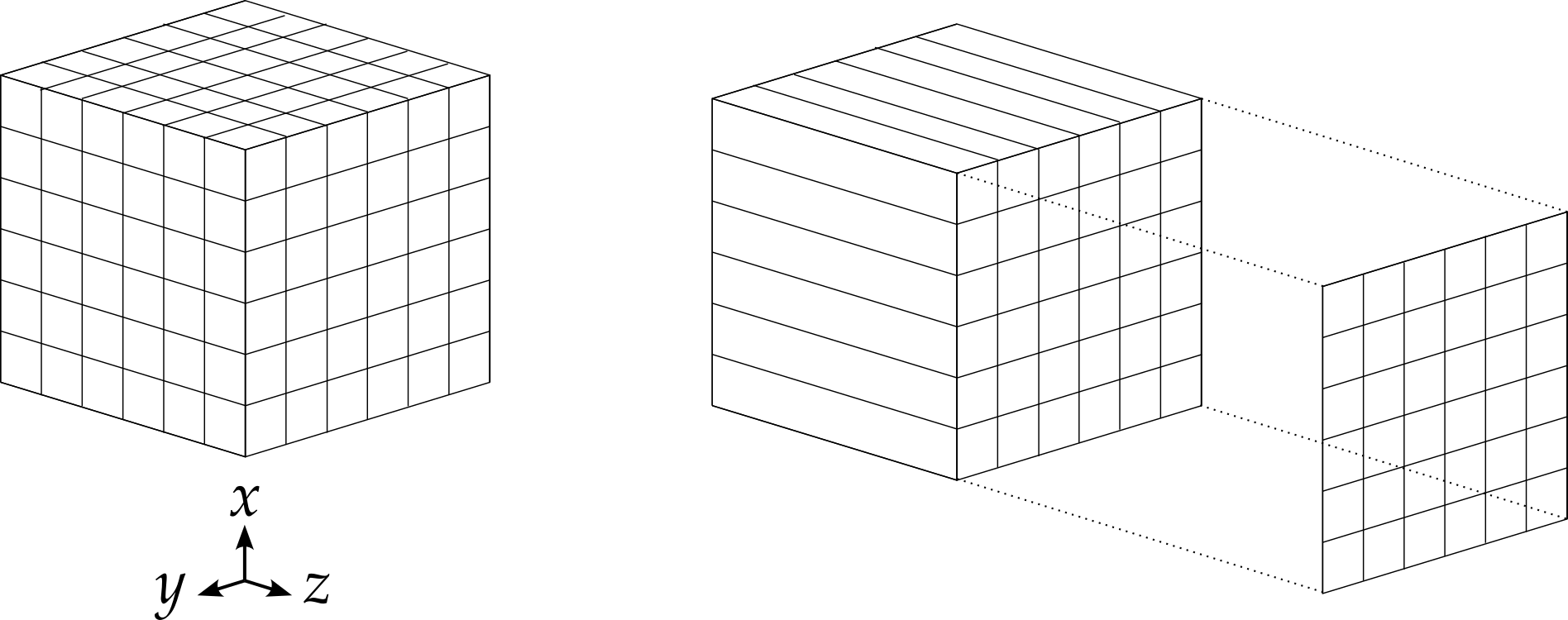}
    \caption{
        \textit{(Left.)} Graphical representation of a 3D variable as a 3D \textsc{PyTorch} tensor.
        \textit{(Right.)} Graphical representation of how radiative transfer is performed, by solving the line integrals along the $z$-axis of the tensor (reminiscent of ``pommes frittes''), producing an image in the $xy$-plane.
    }
    \label{fig:tensor_model}
\end{figure}

In \textsc{pomme}, every model variable, such as abundance, velocity components, and temperature, is represented by a \textsc{PyTorch} tensor, such that 1D, 2D, or 3D spatial distributions are represented by 1D, 2D, or 3D tensors respectively, as implemented in the class \textsc{TensorModel} in \textsc{pomme} (see e.g.\ Figure \ref{fig:tensor_model}, left hand side).
These tensors are viewed as regular Cartesian grids containing the values of the variables and are referred to as ``fields''.
Model variables that are not described by a distribution, such as the mass of a central object, are represented by 1D, single value, tensors.
All model variables can either be ``free'' or ``fixed'', depending on whether they should be optimised (and thus updated) in the reconstruction process.
This property directly maps to the ``requires\_grad'' property of \textsc{PyTorch} tensors.
Furthermore, several convenience functions are provided to easily create astrophysical models, for instance, to create spherically symmetric models or to map point cloud data to a Cartesian grid.
These are demonstrated in the applications in Sections \ref{sec:tests} and \ref{sec:applications}, and the complete \textsc{jupyter} notebooks can be found in the online documentation\footnote{The \textsc{pomme} documentation can be found at: \url{https://pomme.readthedocs.io}.}.

In \textsc{pomme}, the line integrals of radiative transfer (Eqs.\ \ref{eq:formalsolution} and \ref{eq:opticaldepth}) can currently only be performed along an axis of the Cartesian grid, i.e.\ along an axis of the tensors representing the model variables\footnote{For computational efficiency, these reductions are preferably performed along the last axis, since, for that axis, the data is contiguous in memory.} (see e.g.\ Figure \ref{fig:tensor_model}, right hand side).
In this way, integrals can be implemented in \textsc{PyTorch} using the efficient sum operations (reductions) along the tensor axis.
The numerical details of the implementation are discussed in Appendix \ref{app:impl}.
This design choice poses a problem when the radiation field needs to be computed in a direction that does not align with one of the coordinate axis, for instance, when an angle-averaged mean intensity is required.
This can be solved by rotating the model within the model box, as we will show in future work.

To solve the optimisation problem of maximising the log posterior, or minimising the loss, as described in Section \ref{subsec:inverse}, for the examples in this paper, we used the \textsc{Adam} algorithm \citep{kingma_adam_2015}, as implemented in \textsc{PyTorch}.
However, in general, any other solution method implemented in \textsc{PyTorch} can readily be used in \textsc{pomme}.
The particular optimisation schemes, for instance, the learning rate and its variation over the iterations, are very problem dependent, and can be found in the online documentation for the problems presented in this paper.

The computation of visibilities (Eq.\ \ref{eq:visibilities}) in \textsc{pomme} is based on \textsc{Galario}\footnote{The source code can be found at: \url{https://github.com/mtazzari/galario}.}, and, in particular, on the \textsc{numpy} implementation that they use for testing, presented in \cite{tazzari_galario_2018}, which can readily be translated from the \textsc{numpy} framework into \textsc{PyTorch}.

\subsection{Using \textsc{pomme}}
\textsc{pomme} can be a powerful modelling tool but it should be used with great care.
We cannot yet simply feed any spectral line observation into \textsc{pomme} and always get a reasonable reconstruction.
In the following sections, we discuss the crucial choices that carefully have to be made to obtain a decent result.

\subsubsection{Initialisation}
The performance of optimisation methods based on gradient descent critically depends on the initialisation of the underlying iterative procedure. This is especially the case for high-dimensional problems with many optima in the loss function, like the ones considered here.
Choosing an unfortunate initial model can cause severe over-fitting of some parameters while neglecting others, or it might impede minimisation of the loss altogether.
Initially, we aimed to build a method that could autonomously analyse spectral line observations, but we are not there yet.
The performance of the reconstruction algorithm is still too sensitive to the initial guess, such that human supervision remains crucially important in the process.
For the moment, \textsc{pomme} is thus best used to refine a good first guess of a model, for instance, by making predictions for the deviations from spherical symmetry, as demonstrated with the companion-perturbed stellar wind model in Section \ref{subsec:phantom}, rather than doing the entire analysis autonomously.
\change{
In absence of a good initial guess, one can always be obtained, for instance, by Monte Carlo sampling the parameter space (see also Section \ref{sec:discussion}).
}

\subsubsection{Update step size / learning rate}
Given the initial model, the optimisation algorithm will iteratively update the model parameters in the direction opposite to the gradient of the loss, such that the loss is minimised.
The size of this update depends on what in the machine learning community is known as the learning rate.
In this paper, we used the \textsc{Adam} algorithm \citep{kingma_adam_2015} which adaptively tunes this rate to optimise convergence.
We were able to obtain good results when normalising the total loss, setting the hyper-parameter corresponding to the learning rate\footnote{Note that in \textsc{PyTorch} this parameter is called ``\texttt{lr}'' in the code and ``$\gamma$'' in the documentation, while it is called ``$\alpha$'' in the original paper by \cite{kingma_adam_2015}.} $\alpha=0.1$, while using the \textsc{PyTorch} default values, suggested by \cite{kingma_adam_2015}, for all other internal parameters.

\subsubsection{Relative weights of the different loss functions}
The relative weights of the different loss functions can be used as a way to quantify the relative uncertainties in the constraints that they represent.
As such, they are usually fixed hyper-parameters of the optimisation problem.
We found, however, that the performance of the optimisation procedure could be enhanced by monitoring the individual losses and enhancing the weight of a loss once its improvement started to stagnate.
In this way, we can manually tweak the focus of the optimisation process.
In future work, we will study the precise impact of this on the interpretation of the posterior distribution.
\change{
The particular choices that were made in the examples presented in this paper can be found in the online documentation.
}

\section{Proof of concept}
\label{sec:tests}
To demonstrate the possibilities of our reconstruction method, we apply it to two sets of synthetic observations of stellar wind models.
For simplicity, we start with a spherically symmetric model, and then consider an intricate 3D hydrodynamics model of a companion-perturbed stellar wind.
In both cases, the synthetic observations were obtained with the forward model implemented in \textsc{pomme}, so we only test the reconstruction capabilities here.
The forward model itself was benchmarked before against the \textsc{Magritte} line radiative transfer solver \citep{de_ceuster_3d_2022}.

\subsection{Spherically symmetric stellar wind model}
First, we consider a simple spherically symmetric stellar wind model. The model box is given by the radial coordinate $r \in [r_{\star}, r_{\text{out}}] = [1, 10^{4}] \ \text{au}$.
For the velocity field, we assume a typical radially outward directed $\beta$-law,
\begin{equation}
    v(r) \ = \ v_{\star} \ + \ \left( v_{\infty} - v_{\star} \right) \left(1 - \frac{r_{\star}}{r}\right)^{\beta} ,
\label{eq:1D_vel}
\end{equation}
in which $v_{0} = 0.1 \ \text{km}/ \text{s}$,  $v_{\infty} = 20 \ \text{km}/ \text{s}$, and $\beta=0.5$.
We assume the density and velocity to be related through the conservation of mass, such that,
\begin{equation}
    \rho \left( r \right) \ = \ \frac{\dot{M}}{4 \pi r^{2} \, v(r)},
\label{eq:spherical_cont}
\end{equation}
where, for the mass-loss rate, we take a typical value of $\dot{M} = 5.0 \times 10^{-6} \ M_{\sun} / \text{yr}$.
The CO abundance is assumed to be proportional to the density, such that, $n^{\text{CO}}(r) = 3.0 \times 10^{-4} \, N_{A} \, \rho(r) / m^{\text{H}_2}$, with $N_{A}$ Avogadro's number, and $m^{\text{H}_2} = 2.02 \ \text{g}/\text{mol}$, the molar mass of $\text{H}_{2}$.
For the gas temperature, we assume a power law,
\begin{equation}
    T(r) \ = \ T_{\star} \left(\frac{r_{\star}}{r}\right)^{\epsilon} ,
\label{eq:1D_temp}
\end{equation}
with $T_{\star} = 2500 \ \text{K}$, and $\epsilon=0.6$.
Finally, we assume a constant turbulent velocity $v_{\text{turb}}(r) = 1 \ \text{km}/\text{s}$.
The model is discretised on a logarithmically-spaced grid consisting of 1024 elements with $r \in [r_{\text{in}}, r_{\text{out}}] = [10^{-1}, 10^{4}] \ \text{au}$.
Note that $r_{\text{in}} < r_{\star}$, such that several rays hit the stellar surface, since, for concenience, we use the same discretisation for the impact parameters of the rays. 
We impose a boundary condition at $r=r_{\star}$, such that the part of the model inside the star ($r<r_{\star}$) does not contribute to the resulting observation, and thus cannot be reconstructed.

We base our reconstructions on synthetic observations of two commonly observed rotational CO lines $J = \{(3-2), \, (7-6)\}$, which we observe, each in 50 frequency bins, centred around the lines, with a spacing of 1.02 km/s.
\change{
A summary of all reconstructions in this paper can be found in Table \ref{tab:sum} in Appendix \ref{app:figures}.
}

\subsubsection{Reconstructing only the mass-loss rate}
\label{subsubsec:Mdot}
As a first test, we aim to reconstruct only the mass-loss rate.
We initialise the model parameters with all the correct values and only set the mass-loss rate with ten different initial values $\dot{M} = \{1, 2, ..., 10 \} \times 10^{-6} \ M_{\sun} / \text{yr}$.
For all of these initial models, we aim to fit the two CO lines, using the mass-loss rate, $\dot{M}$, but also $v_{\star}$, $v_{\infty}$, $\beta$, $T_{\star}$, and $\epsilon$, as free parameters.
Since we initialised all parameters except the mass-loss rate with the exact value of the original model that we aim to reconstruct, it might seem odd to fit these again.
However, we intentionally do this to demonstrate the limitations of the algorithm, even when parameters are initialised with exactly the ``right'' values.
For each initialisation, we run the optimisation algorithm for 500 iterations, minimising the relative differences between the synthetic observations well below $10^{-3}$, i.e.\ below the noise level of any realistic observation.

Table \ref{tab:Mdot_rec_param} shows the resulting reconstructed parameters.
Overall, the mass-loss rates are properly retrieved and also the other parameters do not deviate significantly from their ``right'' values.
The second and third model seem to have gotten stuck in a local (sub-)optimum that is slightly away from the ``right'' values.
This illustrates that results depend on the initialisation of the reconstruction process, caused by the lack of information in the observations.

\begin{deluxetable}{r | c c c c c c c c}
\tablecaption{Reconstructed model parameters, based on synthetic observations of the CO $J = \{(3-2), \, (7-6)\}$ lines, for a spherically symmetric stellar wind model, initialised with different mass-loss rates (see Section \ref{subsubsec:Mdot}).}
\label{tab:Mdot_rec_param}
\tablehead{ $\dot{M}_{\text{ini}}$ [$M_{\sun}$/yr]  &  $\dot{M}_{\text{rec}}$ [$M_{\sun}$/yr]  &  $v_{\star}$ [km/s]  &  $v_{\infty}$ [km/s]  &  $\beta$  &  $T_{\star}$ [K]  &  $\epsilon$ }
\startdata
$1.0 \times 10^{-6} $  &  $5.0 \times 10^{-6}$  &  0.09  &  20.00  & 0.51  &  2499  &  0.60 \\
$2.0 \times 10^{-6} $  &  $4.8 \times 10^{-6}$  &  0.06  &  19.97  & 1.12  &  2350  &  0.59 \\
$3.0 \times 10^{-6} $  &  $4.9 \times 10^{-6}$  &  0.07  &  20.00  & 0.67  &  2468  &  0.60 \\
$4.0 \times 10^{-6} $  &  $5.0 \times 10^{-6}$  &  0.09  &  20.00  & 0.55  &  2491  &  0.60 \\
$5.0 \times 10^{-6} $  &  $5.0 \times 10^{-6}$  &  0.10  &  20.00  & 0.50  &  2500  &  0.60 \\
$6.0 \times 10^{-6} $  &  $5.0 \times 10^{-6}$  &  0.09  &  20.00  & 0.50  &  2499  &  0.60 \\
$7.0 \times 10^{-6} $  &  $5.0 \times 10^{-6}$  &  0.10  &  20.00  & 0.50  &  2500  &  0.60 \\
$8.0 \times 10^{-6} $  &  $5.0 \times 10^{-6}$  &  0.10  &  20.00  & 0.53  &  2498  &  0.60 \\
$9.0 \times 10^{-6} $  &  $5.0 \times 10^{-6}$  &  0.10  &  20.00  & 0.47  &  2501  &  0.60 \\
$1.0 \times 10^{-5} $  &  $5.0 \times 10^{-6}$  &  0.10  &  20.00  & 0.51  &  2493  &  0.60 \\
\enddata
\end{deluxetable}

\subsubsection{Reconstructing only the CO distribution}
\label{subsubsec:only_CO}
Next, we aim to reconstruct only the CO abundance distribution, while keeping the velocity and temperature fixed at their correct distributions.
We start from a simple quadratically decreasing initial CO distribution, $n_{\text{CO}}(r) = 5.0 \times 10^{14} \, \text{m}^{-3} \, (r_{\text{in}}/r)^{2} $, where the prefactor is chosen such that the resulting synthetic observations of the lines roughly match with those of the original model.
As regularisation, we first impose no prior, then only a smoothness prior on the CO abundance distribution (Eq.\ \ref{eq:reg_smooth}), and finally also a second regularisation term that enforces the continuity equation (\ref{eq:3D_cont}), which, in spherical coordinates, reads,
\begin{equation}
    \mathcal{L}[\rho, v]
    \ = \
    \int_{0}^{\infty} 4\pi r^{2} \text{d}r \left\{ \frac{1}{\rho \, r^{2}} \, \partial_{r} \left( r^{2} \rho \, v \right) \right\}^{2} .
\label{eq:1D_cont}
\end{equation}
We run the optimisation algorithm for 500 iterations, minimising the relative differences between the synthetic observations well below $10^{-3}$, i.e.\ below the noise level of any realistic observation.

\begin{figure}
    \centering
    \includegraphics[width=\linewidth]{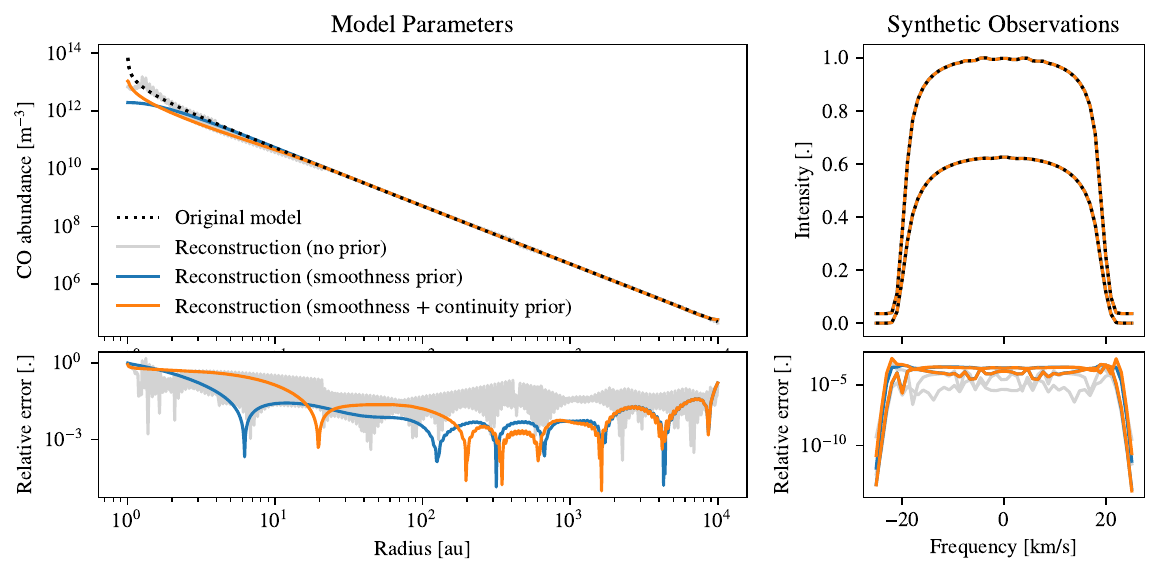}
    \caption{
        Reconstructions of the CO abundance distribution, synthetic observations of the CO $J = \{(3-2), \, (7-6)\}$ lines, and the absolute relative differences between the reconstructions and the models, and between their corresponding synthetic observations.
    }
    \label{fig:1D_only_CO}
\end{figure}

The resulting reconstructions and their corresponding synthetic observations are shown in Figure \ref{fig:1D_only_CO}.
The synthetic observations are all clearly consistent with the original model.
The reconstructions of the CO abundance distribution are good with only small deviations close to the star.
This is to be expected, since the region close to the star is obscured most by other material such that its contribution to the resulting observations is convoluted the most.
The reconstruction without any prior is not very smooth, due to the lack of a smoothness prior.
The reconstruction that only used the smoothness prior but not the continuity prior, is much smoother, but shows a different trend towards the star, missing the sharp increase in density.
This could wrongfully be interpreted as a sudden decrease in mass-loss rate.
Including the continuity prior prevents this different trend, tying the density to the velocity field, although the overall difference with the original model is slightly larger.

\subsubsection{Reconstructing the CO, temperature, \& velocity distribution}
\label{subsubsec:all}
Finally, we aim to reconstruct the CO abundance, the temperature, and the velocity distributions.
As in the previous example, we start from a quadratically decreasing initial CO distribution, $n_{\text{CO}}(r) = 5.0 \times 10^{14} \, \text{m}^{-3} \, (r_{\text{in}}/r)^{2} $, where the prefactor is chosen such that the resulting synthetic observations of the lines roughly match with those of the original model.
As regularisation, we first impose no prior, then only a smoothness prior on the three distributions (Eq.\ \ref{eq:reg_smooth}), and finally also a second regularisation term that enforces the continuity equation (\ref{eq:3D_cont}), for spherically symmetric models given by equation (\ref{eq:1D_cont}).
We run the optimisation algorithm for 500 iterations, minimising the relative differences between the synthetic observations well below $10^{-3}$, i.e.\ below the noise level of any realistic observation.

\begin{figure}
    \centering
    \includegraphics[width=\linewidth]{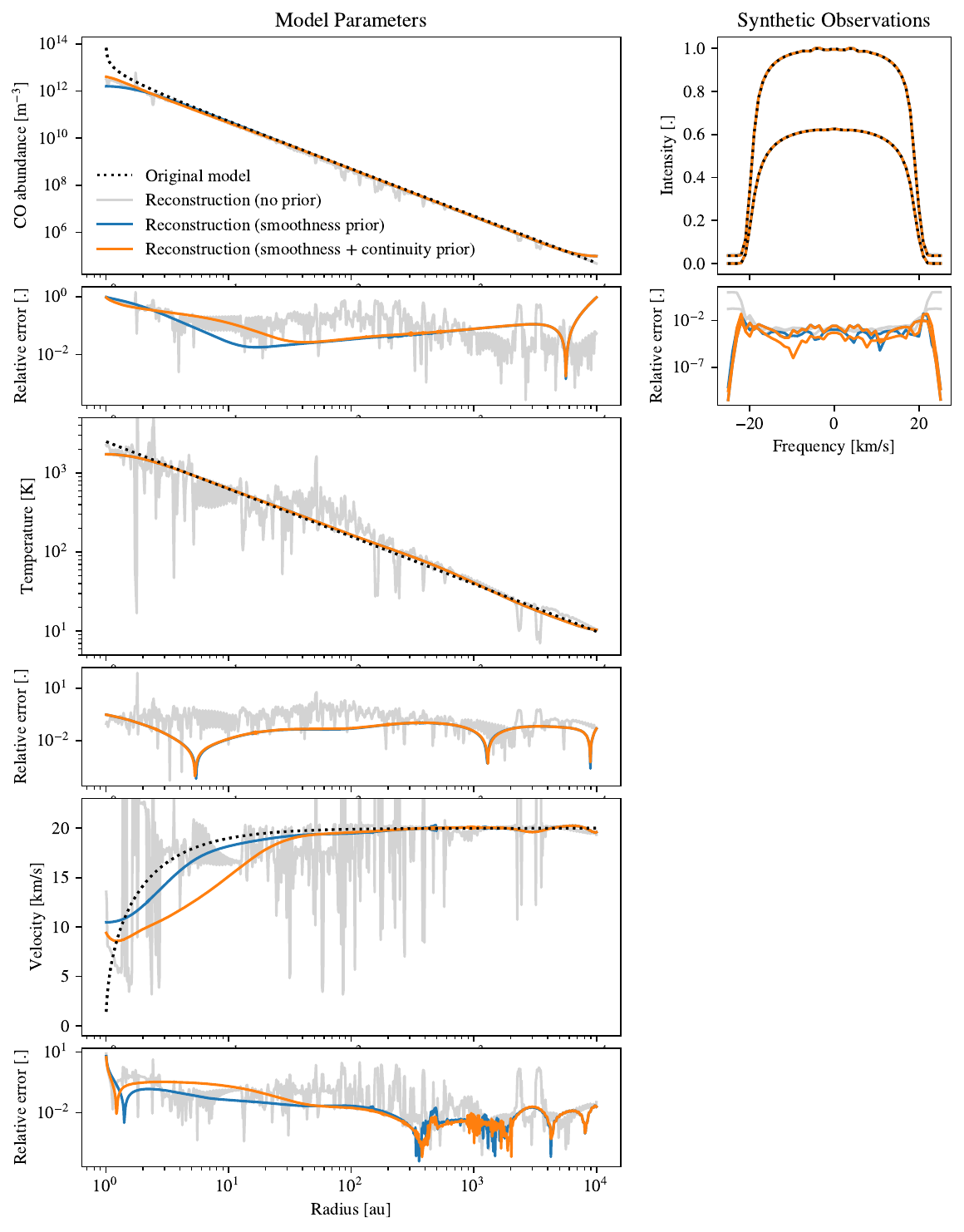}
    \caption{
        Reconstructions of the CO abundance, temperature, and velocity distributions, synthetic observations of the CO $J = \{(3-2), \, (7-6)\}$ lines, and the absolute relative errors between the reconstructions and the models, and between their corresponding synthetic observations.
    }
    \label{fig:1D_all}
\end{figure}

Figure \ref{fig:1D_all} shows the reconstructions and their corresponding synthetic observations.
The synthetic observations are clearly consistent with the observations of the original model.
The reconstruction without any prior is very irregular, due to the lack of smoothness prior, and only very roughly follows the trends of the original model distributions.
The other two reconstructed models (with smoothness prior), however, show deviations from the original model, especially the velocity distribution, but only close to the star.
The difficulty to reconstruct the region close to the star can again be understood, since that region is obscured the most by other material and thus its contribution to the observations is convoluted the most.
The difficulty to reconstruct the velocity near the star, moreover, is a result of the degeneracy between the model parameters.
The different CO abundance and velocity distributions compared to the original model yield almost exactly the same observations, such that the algorithm cannot improve much on this result.
Additional observations or stricter prior assumptions could break this degeneracy.

\subsection{Companion-perturbed 3D stellar wind model}
\label{subsec:phantom}
\begin{figure}
    \centering
    \includegraphics[width=\linewidth]{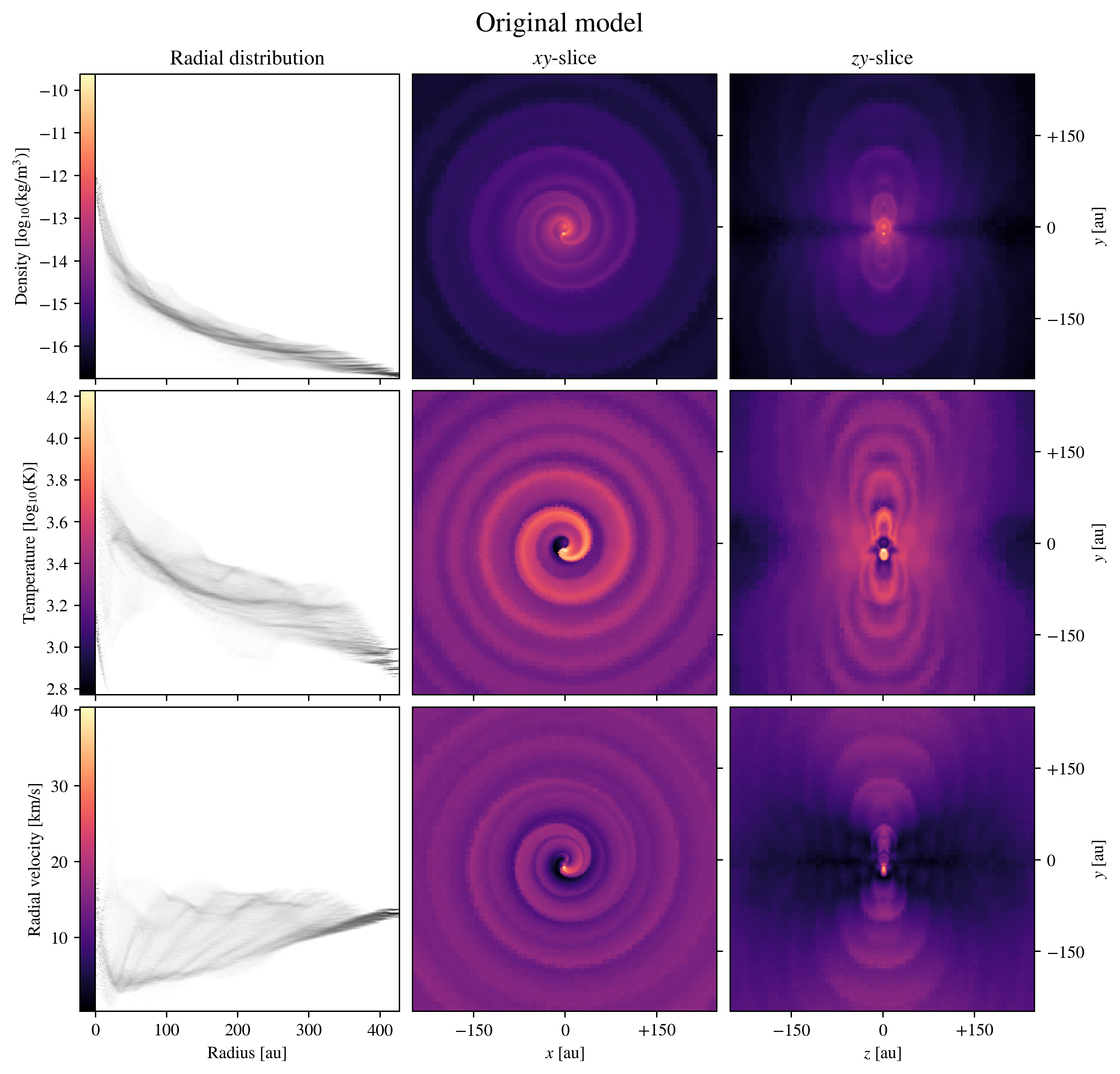}
    \caption{
        Radial distribution and two slices of the original model of the density, temperature, and radial velocity distributions of the companion-perturbed AGB wind hydrodynamics model that we aim to reconstruct. The radial distributions are obtained with 2D histograms with $428^{2}$ bins, evenly spaced within the parameter range of the model. The colour bars also represent the vertical axis of the radial distributions.
    }
    \label{fig:phantom}
\end{figure}

Next, we consider a much more intricate reconstruction problem based on synthetic observations of a smoothed-particle hydrodynamics (SPH) model of a companion-perturbed stellar wind model provided by J.\ Malfait \citep[see][\color{black} Malfait et al.\ \textit{subm.}]{malfait_sph_2021, maes_sph_2021, siess_3d_2022, esseldeurs_3d_2023}\footnote{In contrast to earlier work \citep[see e.g.][]{malfait_sph_2021, maes_sph_2021}, the SPH model includes HI-cooling, resulting in (more) realistic temperature distributions (see also Malfait et al.\ \textit{subm.}).}.
The density, temperature and radial velocity field of the SPH model are displayed in Figure \ref{fig:phantom}, where we see the intricate spiral structure etched in these distributions by the companion.
We base our reconstruction on synthetic observations along the $z$-axis, i.e.\ the $xy$-plane corresponds to the plane of the sky and we view the spiral structure face-on. We use two commonly observed rotational lines, CO ($J=4-3$) and SiO ($J=3-2$), each observed in 100 frequency bins, centred around the lines, with a spacing of $0.12$ km/s.
The synthetic observations are shown in Figures \ref{fig:phantom_obs_CO} and \ref{fig:phantom_obs_SiO} in Appendix \ref{app:figures}.
As starting point for the reconstruction process, we use a spherically symmetric model, for which the parameters are summarised in Table \ref{tab:init_param}. Initially, we assume a density distribution of the form, $n_{\text{H}_{2}}(r) = n^{\text{H}_{2}}_{\star} \left(r_{\star} / r\right)^{2}$, a radial $\beta$-type velocity field (Eq. \ref{eq:1D_vel}), and a power-law temperature distribution (Eq. \ref{eq:1D_temp}). Throughout the reconstruction, we assume constant abundances of CO and SiO relative to H$_2$, i.e.\ we only reconstruct the assumed common underlying density distribution.
The specific initial parameters are chosen such that the integrated spectral energy distributions of the synthetic observations roughly match with those of the original model.
As regularisation, we first impose no prior, then only a smoothness prior for the density, temperature, and radial velocity distributions, given by equation (\ref{eq:reg_smooth}), and finally, we also add a regularisation term that enforces continuity, given by equation (\ref{eq:3D_cont}).
For the reproduction loss, we normalised the contributions of every line and pixel with the frequency-integrated results (see also Eq.\ \ref{eq:relative_loss}). However, to reduce the dynamical range of the frequency-integrated results, we applied a logarithm before computing the differences, such that the resulting reproduction loss yields a slight deviation from equation (\ref{eq:relative_loss}),
\begin{equation}
    \mathcal{L}_{\text{rep}}\big(f(\boldsymbol{m}), \boldsymbol{o} \big)
    \ = \
    \Big \|\log \big\langle f(\boldsymbol{m}) \big\rangle - \log \left\langle\boldsymbol{o}\right\rangle \Big\|^{2}
    \ + \
    \left \| \frac{f(\boldsymbol{m})}{\big\langle f(\boldsymbol{m})\big\rangle} - \frac{\boldsymbol{o}}{\left\langle \boldsymbol{o}\right\rangle} \right\|^{2} ,
\end{equation}
where the brackets, $\langle\cdot\rangle$, denote the integral over frequency, and the bars, $\|\cdot\|$, denote the Euclidean norm.
Starting from this initial model, we reconstructed the density, temperature, and (purely) radial velocity distributions, in a $64^{3}$-element model box, while all other parameters were set to their correct values.

\begin{deluxetable}{c c c c c c c c c}
\tablecaption{Parameters for the initial model in the reconstruction of the companion-perturbed 3D stellar wind.}
\label{tab:init_param}
\tablehead{$r_{\star}$ [au]  &  $n^{\text{H}_{2}}_{\star}$ [m$^{-3}$]  &  $n^{\text{CO}}$/$n^{\text{H}_{2}}$  &  $n^{\text{SiO}}$/$n^{\text{H}_{2}}$  &  $v_{\star}$ [km/s]  &  $v_{\infty}$ [km/s]  &  $\beta$  &  $T_{\star}$ [K]  &  $\epsilon$ }
\startdata
6.72  &  $5.0 \times 10^{13}$  &  $3.0 \times 10^{-4}$  &  $5.0 \times 10^{-6}$  &  0.1  &  10  &  0.5  &  5000  &  0.5
\enddata
\end{deluxetable}

\begin{figure}
    \centering
    \includegraphics[width=\linewidth]{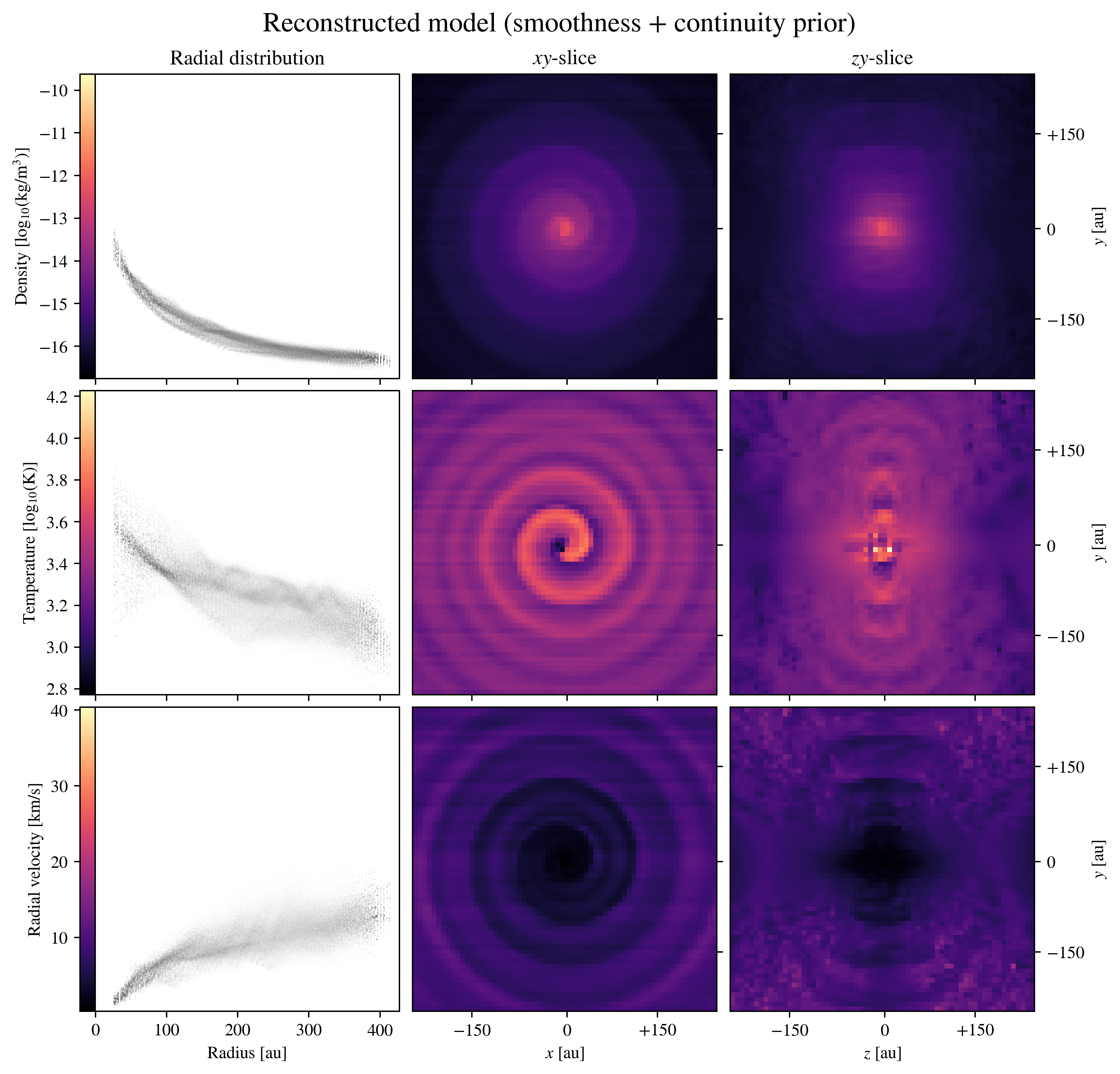}
    \caption{
        Radial distribution and two slices of the reconstruction (with smoothness and continuity prior) of the density, temperature, and radial velocity distributions of the companion-perturbed AGB wind hydrodynamics model. The radial distributions are obtained with 2D histograms with $428^{2}$ bins, evenly spaced within the parameter range of the original model. The colour bars also represent the vertical axis of the radial distributions.
    }
    \label{fig:phantom_rec}
\end{figure}

\begin{figure}
    \centering
    \includegraphics[width=0.85\linewidth]{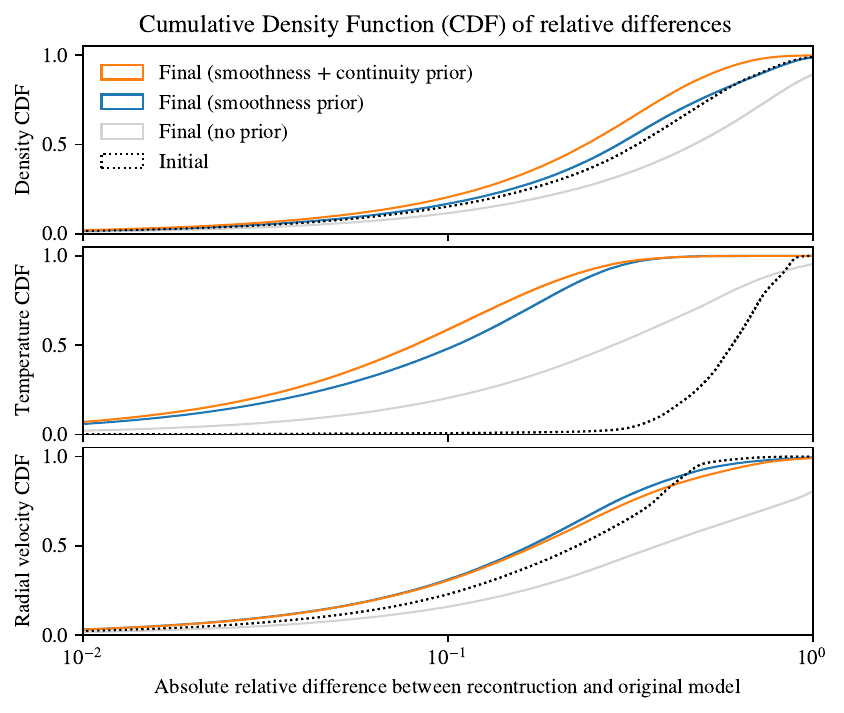}
    \caption{
        Cumulative distribution functions (CDFs) of the absolute relative differences between the reconstructions and the original model.
        A faster rising curve implies more smaller differences and thus a better reconstruction.
    }
    \label{fig:phantom_err}
\end{figure}

\change{
Figure \ref{fig:phantom_rec} shows the reconstructed model using both the smoothness and continuity prior of the stellar wind model after 600 iterations.
The corresponding reconstructions using an equivalent setup, but using no prior and only using the smoothness prior are shown in Figures \ref{fig:phantom_rec_nono} and \ref{fig:phantom_rec_noco} in Appendix \ref{app:figures} respectively.
The reconstruction without any prior (Fig. \ref{fig:phantom_rec_nono}) clearly lacks regularity, due to the absence of the smoothness prior.
The reconstruction only using the smoothness prior (Fig. \ref{fig:phantom_rec_noco}) shows no spiral structure in the velocity field, due to the missing connection between density and velocity from the continuity prior.
The reconstruction using both smoothness and continuity prior (Fig. \ref{fig:phantom_rec}), qualitatively bears quite a close resemblance to the original model, although, overall, features appear less sharp in the reconstruction.
The greatest differences appear in the radial velocity distribution.
This can be understood, since, in the original model the velocity field is not purely radial, while we assume it to be purely radial in the reconstruction.
Even though we used observations of the model along the $z$-axis (and thus in the $xy$-plane), also the orthogonal $zy$-plane seems to be reconstructed reasonably well.
}

\change{
Figure \ref{fig:phantom_err} shows a quantitative assessment of the reconstruction quality, with the cumulative distribution functions of the absolute relative differences\footnote{If $\boldsymbol{m}_{\text{ori}}$ and $\boldsymbol{m}_\text{rec}$ are vectors containing the original and reconstructed model parameters respectively, then we define the vector of absolute relative differences as $2 | \boldsymbol{m}_{\text{ori}} - \boldsymbol{m}_{\text{rec}} | / ( \boldsymbol{m}_{\text{ori}} + \boldsymbol{m}_{\text{rec}} )$, in which all operations are element-wise.} between the original and reconstructed model parameters.
Although there are clear differences between the reconstruction and the original model, we see a clear improvement with respect to the initial (spherically symmetric) guess for the model, and a further improvement by adding the priors.
Also quantitatively, the radial velocity performs the worst, with even a slight increase in large differences with respect to the initial model, while the temperature shows the biggest improvement, with half of the absolute relative differences below 10\%.
The reconstruction without any prior  for the density and radial velocity is considerably worse than the initial guess, further emphasising the importance of the smoothness prior as a minimal assumption.
}

\section{Application}
\label{sec:applications}
To demonstrate the utility of our methods, we apply it to a reconstruction problem from the literature, using real (in contrast to synthetic) astronomical observations.

\subsection{The NaCl distribution around the AGB star IK Tau}
In their recent paper, \cite{coenegrachts_unusual_2023} studied the unusual distribution of NaCl around the oxygen-rich AGB star IK Tau. They did this by deprojecting the spectral line emission detected with the Atacama Large (sub)Millimetre Array (ALMA), as such creating a 3D model of the NaCl distribution.
This was done by assuming a spherically symmetric, radially outward-directed, $\beta$-type (Eq. \ref{eq:1D_vel}) velocity field, centred around the star.
In that way, for every viewing angle, the projected velocity along the line of sight is an invertible function.
If, then, all observed emission is assumed to originate from the line centre, any frequency shift in the observed emission can be attributed to Doppler shifts caused by motion along the line of sight. Since this motion has an invertible velocity field (by assumption), the observed emission at each frequency can be associated with a unique position along the line of sight and a 3D emission map can be obtained.
Using an empirical relation between the line emission and the abundance, a 3D distribution of the NaCl abundance can be obtained.
It should be noted that up to this point only the assumed velocity structure together with the observation determine the abundance distribution.
\cite{coenegrachts_unusual_2023} then further refined this estimate for the abundance distribution by introducing scaling factors for the abundance in five different clumps that appeared in the model and fitting these factors by comparing the synthetic observations from a radiative transfer model with the real observations.

Although this approach gives a good first estimate of the 3D distribution of NaCl, there are some shortcomings.
First of all, there is the strong assumption on the velocity field, a $\beta$-law, which only has three free parameters.
Second, there is the limitation that scaling factors for the abundance can only be fitted for a few structures and cannot account for any variation of the abundance within those structures.
As with most fitting tools, the number of free parameters in the model is rather limited.
With \textsc{pomme}, this modelling can be done significantly faster and with many more free parameters, giving a more realistic prediction for the NaCl distribution.

\cite{coenegrachts_unusual_2023} created their deprojected model from synthesised images that map the intensity in the plane of the sky, based on interferometric (ALMA) data.
Therefore, we also start from the same synthesized channel maps.
They used \textsc{CASA} \citep{the_casa_team_casa_2022} to add the instrumentation effects to their synthetic observations.
As explained in Section \ref{subsubsec:observationalinstrumentationeffects}, we only convolve our synthetic images with the beam associated with the synthesised channel maps.

\subsubsection{Reconstructing the NaCl distribution}
\label{subsubsec:rec_NaCl}
We start by assuming the same spherically symmetric temperature distribution and velocity field as \cite{coenegrachts_unusual_2023}, and we only reconstruct the NaCl distribution using \textsc{pomme}.
As initial guess for the NaCl distribution, we take a spherically symmetric distribution centred around the star, given by,
\begin{equation}
    n_{\text{NaCl}}(r) = 10^{10} \ \text{m}^{-3} \left(\frac{5.34 \ \text{au}}{r}\right)^{2} ,
\end{equation}
where $r$ is the distance from the star in the model.
As regularisation, we only impose a smoothness prior on the NaCl abundance of the form (Eq. \ref{eq:reg_smooth}).
We do not impose a hydrodynamics prior, since the underlying dynamics is expected to be far from a steady state, and hence, unsurprisingly, no choice of parameters in the steady-state hydrodynamic prior improved the reconstruction.

\begin{figure}
    \centering
    \includegraphics[width=\linewidth]{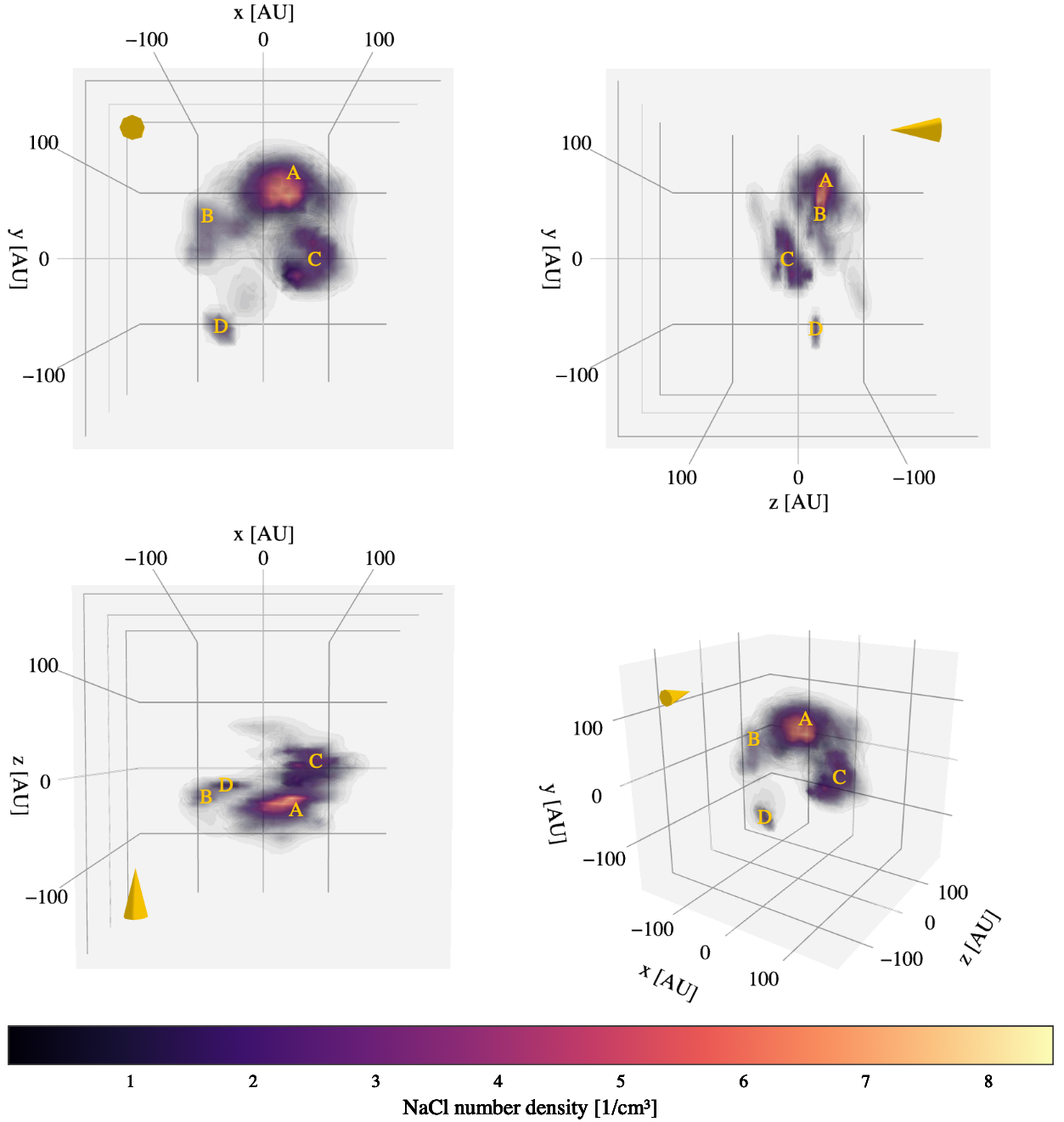}
    \caption{
        3D rendering of the reconstructed NaCl distribution around IK Tau, viewed from different angles.
        The yellow cone indicates the viewing direction of the observations. An interactive version of this figure can be found in the online documentation for \textsc{pomme} at \url{https://pomme.readthedocs.io/en/latest/_static/NaCl_reconstruction.html}.
    }
    \label{fig:IKTau_NaCl_3D}
\end{figure}

Figure \ref{fig:IKTau_NaCl_3D} shows a 3D rendering of the resulting reconstructed NaCl distribution, obtained with \textsc{pomme}, viewed from different angles.
Overall the distribution looks very similar to the one obtained by \cite{coenegrachts_unusual_2023}, compare e.g.\ with their Figure 4.
However, note that their 3D renderings show 3D emission maps, whereas our Figure \ref{fig:IKTau_NaCl_3D} shows the actual abundance distribution.
Figure \ref{fig:IKTau_obs} in Appendix \ref{app:figures} shows the observations and the relative difference between the synthetic observations of the reconstructed model and the true observations.
We will not repeat the entire analysis of this observed emission here, since our findings are very similar, and we only aim to demonstrate the capabilities of \textsc{pomme}.
One thing to point out are the finer structural details that can be obtained with \textsc{pomme}.
For instance, the clump that \cite{coenegrachts_unusual_2023} identified as ``clump C" in their analysis shows in our reconstruction much more structure. The clump takes the shape of a small arc, suggesting a spiral, originating from the star, which further supports the hypothesis of the formation by binary interaction.
\change{The reason we find this finer structure is that we do not require any interpolation between different model meshes as \cite{coenegrachts_unusual_2023} did.}
Another important point is that while the modelling pipeline of \cite{coenegrachts_unusual_2023}, including deprojection and LTE radiative transfer with \textsc{Magritte} \citep{de_ceuster_3d_2022}, takes several minutes per iteration, in our implementation with \textsc{pomme}, the entire reconstruction with 500 iterations can be done in less than three minutes.
This speed-up is mainly because in \textsc{pomme} all calculations can be performed on the tensor representation of the model, avoiding computationally costly interpolations to another radiative transfer model.

\section{Discussion}
\label{sec:discussion}

\subsection{Comparison to literature}
Given the vast literature on reconstruction methods and the specific applications to stellar wind modelling we have in mind, we will focus our discussion on reconstruction methods that have been used in the literature on evolved stars.

It should be emphasised that, in contrast to what in the literature is often termed deprojection methods \citep[see e.g.][]{guelin_irc_2018, montarges_noema_2019}, \textsc{pomme} does not merely deproject the observed radiation into an emission map, but can directly reconstruct model parameters, such as density, velocity, or temperature distributions that caused the observed radiation.

Furthermore, a key property of \textsc{pomme} is that the forward model is implemented in \textsc{PyTorch}, such that we can leverage the \textsc{autograd} functionality \citep{paszke_automatic_2017} and conveniently compute gradients through the model.
This allows us to use gradient descent based methods to fit the model to observations.
In contrast, for instance, \cite{coenegrachts_unusual_2023} rely on an external radiative transfer solver such that fitting parameters to observations requires less efficient Monte Carlo searches through the often vast parameter space.
This, in combination with the optimised implementation in \textsc{PyTorch}, and the fact that it avoids interpolations between different frameworks, makes \textsc{pomme} a computationally efficient tool to create model reconstructions (see also Section \ref{sec:applications}).

\change{
It should be emphasised, though, that the computational efficiency in \textsc{pomme} comes at the cost of knowledge about the full posterior distribution.
In {\sc pomme}, we aim to find the mode (i.e.\ maximum) of the posterior as fast as possible, this means, with the least number of evaluations of the forward model, or correspondingly, with the least number of samples from the posterior distribution.
As a result, we are sensitive to the initial model guess and have difficulty with model degeneracy.
Therefore, {\sc pomme} works best, when one already has an idea of what the resulting reconstruction should look like.
Otherwise, an extensive Monte Carlo search could be used to more thoroughly explore the parameter space and posterior distribution, but this comes at a significant computational cost.
}

For relatively simple structures, such as discs or spirals, one can often fit analytic models to observations using only a few parameters
\citep[see e.g.][]{mauron_imaging_2006, bitsch_structure_2015, homan_alma_2018, homan_unusual_2018}.
This yields simple models that are relatively easy to fit and interpret.
Such analytic models can also be implemented in \textsc{pomme} to benifit from the efficient implementation of the forward model and the fitting procedure.
However, what is perhaps more interesting, is that, in the Bayesian framework of \textsc{pomme}, these analytic models can also be used as a prior while fitting the (unconstrained) density, temperature and velocity distributions.
This is an alternative approach to the priors based on steady-state hydrodynamics discussed in Section \ref{subsubsec:prior},
and would allow one to fit observations beyond the limits of the idealised analytic models, while keeping relatively strict control over the structure of the reconstructed distributions.
It would be interesting to explore this further, however, we are mainly interested in modelling those observations for which there is no clear analytic structure that might fit the data.
Therefore, we will first focus on the other research avenues discussed below.

\subsection{Current limitations \& future work}
\label{subsec:future}
The true power of the reconstruction method presented in this paper is best showcased by modelling astrophysical objects with highly complex geometries, such as the environments of evolved stars that are almost impossible to model otherwise.
However, due to their complexity, each of these models for specific observations deserves a paper of its own.
Therefore, we have limited ourselves in this paper mostly to merely demonstrating the methodology and we will use this as the starting point for new models of observations in future work.
Furthermore, there are still many ways to extend the work presented here.

First of all, there is the strong assumption of local thermodynamic equilibrium (LTE) in the line formation model that we presented.
This might be valid in dense regions where many collisions between the gas particles drive the medium towards LTE, but is likely not valid in less dense regions where incoming radiation drives the medium out of LTE.
To relax this assumption, a non-LTE radiative transfer solver needs to be implemented, which depends on the mean intensity, and thus requires the radiation field in different directions through the model.
This is a challenge for our approach that heavily relies on efficient operations along tensor axes for line integrals, but we are currently exploring several different ways to overcome this.

Second, there are the many hyper-parameters in our approach, such as the relative weights of the different priors, and the weighting of the prior with respect to the likelihood.
Currently, these all have been tuned by hand for each individual model.
However, since we have sufficiently good models from which we can make synthetic observations to test \textsc{pomme}, knowing what the reconstruction should be (see e.g.\ Section \ref{sec:tests}), we can learn the required hyper-parameters from these examples \citep[see e.g.][]{haber_learning_2003, afkham_learning_2021}.
We are currently exploring how to do this.
The main challenge is to find a method that generalises well enough such that it can be applied to real observations.
In principle, we could go further, and learn the entire prior based on examples, or even learn the entire reconstruction process \citep[see e.g.][]{balakrishnan_visual_2019, diaz_baso_bayesian_2022, ksoll_deep_2023}. 
This, however, requires a lot of faith in the models that are used as training data.
Since we developed our methods with stellar wind applications in mind, for which we are still learning the governing physics and chemistry, especially where companion interactions are concerned, we find ourselves too uncertain about our models to rely on such a data-driven approach.

Third, in this paper, we only considered the mode (i.e.\ the maximum) of the posterior distribution, and identified this with \emph{the} reconstruction of the observation, $\boldsymbol{m}_{\star} = \max_{\boldsymbol{m}} \{ p(\boldsymbol{m} | \boldsymbol{o}) \}$.
Future work could explore the properties of this posterior distribution beyond the mode.
Since we expect this posterior to be quite intricate, it might be that other estimators might be better suited to be identified as \emph{the} reconstruction of an observation, for instance, one could consider the expectation of a model given an observation,
$\mathbb{E}[\boldsymbol{m} | \boldsymbol{o}] \equiv \int \boldsymbol{m} \,  p(\boldsymbol{m} | \boldsymbol{o}) \, \text{d}\boldsymbol{m}$, which could be obtained practically by Monte Carlo sampling from the posterior.
Furthermore, the other moments of the posterior distribution can be used to estimate the goodness of the fit.

\section{Conclusions}
\label{sec:conclusions}
We have presented a Bayesian method to interpret spectral line observations in terms of a 1D or 3D model for physical properties, such as chemical abundance, velocity, and temperature distributions.
In particular, we discussed the implementation of this method in the open-source \textsc{Python} package: \textsc{pomme}.
Furthermore, we presented how prior knowledge, for instance, in the form of a steady-state hydro-dynamics model, can be used to guide the reconstruction process.
We have demonstrated our approach with examples from the stellar wind literature.
As a proof-of-concept, we reconstructed models from synthetic spectral line observations of analytic 1D models and 3D hydrodynamics simulations, and we have shown a first application to real observations by reconstructing the NaCl distribution around the AGB star IK Tau.
We will further showcase the true power of our reconstruction method in a set of forthcoming papers, each dedicated to the particular object we reconstruct.

\begin{acknowledgments}
FDC is a Postdoctoral Research Fellow of the Research Foundation - Flanders (FWO), grant number 1253223N, and was previously supported for this research by a KU Leuven Postdoctoral Mandate (PDM), grant number PDMT2/21/066.
TC is a PhD Fellow of the Research Foundation - Flanders (FWO), grant number 1166722N.
LD acknowledges support from KU Leuven C1 MAESTRO grant C16/17/007, KU Leuven C1 BRAVE grant C16/23/009, KU Leuven Methusalem grant METH24/012, and FWO Research grant G099720N.
TD is supported in part by the Australian Research Council through a Discovery Early Career Researcher Award, number DE230100183, and by the Australian Research Council Centre of Excellence for All Sky Astrophysics in 3 Dimensions (ASTRO 3D), through project number CE170100013.
\end{acknowledgments}

%

\vspace{5mm}
\facilities{ALMA}


\software{
\textsc{astropy} \citep{the_astropy_collaboration_astropy_2013, the_astropy_collaboration_astropy_2018, the_astropy_collaboration_astropy_2022},  
\textsc{PyTorch} \citep{paszke_automatic_2017, paszke_pytorch_2019}
\textsc{Galario} \citep{tazzari_galario_2018}
\textsc{Magritte} \citep[][{\color{black}Ceulemans et al.\ \textit{in prep.}}]{de_ceuster_magritte_2019, de_ceuster_magritte_2020, de_ceuster_3d_2022}
}



\appendix

\section{Numerical implementation of the forward model}
\label{app:impl}

In this appendix, we provide some details about the numerical implementation of the forward model in \textsc{pomme}.

\subsection{Line optical depth}
\label{app:optical_depth}
First, we consider the line optical depth as given in equation (\ref{eq:opticaldepth}).
Velocity gradients play a key role in line radiative transfer.
Since spectral lines are narrowly peaked in frequency space, they are very sensitive to Doppler shifts, and thus motion (gradients), along the line of sight.
Therefore, when numerically solving a line transfer problem, it is key to properly trace the velocity (gradient) along the line of sight. 
Since we assume the line profile to be Gaussian (Eq.\ \ref{eq:profile}), we can take care of this sharp frequency dependence by resolving it analytically {\color{black} (see also Ceulemans et al.\ \textit{subm.})}.

Consider a line-of-sight-segment between two consecutive elements, indexed as 0 and 1, parametrised by $\lambda \in [0, 1]$.
The line optical depth in this segment can then be written as,
\begin{equation}
\chi(\lambda) \ = \ a(\lambda) \, \exp \left(-b(\lambda)^{2}\right) ,
\end{equation}
where we defined,
\begin{align}
a(\lambda) \ &= \ \frac{\chi_{ij}(\lambda) \, n(\lambda)}{\sqrt{\pi} \, \delta\nu_{ij}(\lambda)}, \\
b(\lambda) \ &= \ \frac{1}{\delta \nu_{ij}(\lambda)} \left\{ \left( 1 + \frac{v_{z}(\lambda)}{c} \right) \nu  - \nu_{ij} \right\} .
\end{align}
The narrowly peaked behaviour is mainly caused by the exponential function.
We can resolve this in the computation of the optical depth, for instance, by using linear interpolation functions for $a$ and $b$, while explicitly integrating the exponential.
This yields the optical depth increment\footnote{In the actual implementation we included the factor $\Delta z$ in the definition of $a$, for efficiency.},
\begin{equation}
\Delta \tau \ = \ \Delta z \int_{0}^{1} \text{d}\lambda \ \chi (\lambda) .
\end{equation}
Using a linear interpolation scheme for the functions $a$ and $b$,
\begin{align}
a(\lambda) \ &= \ (1-\lambda) a_{0} \ + \ \lambda a_{1}, \\
b(\lambda) \ &= \ (1-\lambda) b_{0} \ + \ \lambda b_{1},
\end{align}
with discretised values $a_{0}$ and $b_{0}$ at element 0, and with discretised values $a_{1}$ and $b_{1}$ at element 1, this integral yields,
\begin{equation}
\begin{split}
\Delta \tau \ = \ \frac{\Delta z}{2\left(b_{1}-b_{0}\right)^{2}}
    & \Big\{ \left(a_{1}-a_{0}\right) \left( e^{-b_{0}^{2}} - e^{-b_{1}^{2}}\right) \\
    &  \ \ \ + \sqrt{\pi} \left(b_{0}a_{1}-b_{1}a_{0}\right) \big(\text{Erf}(b_{0}) - \text{Erf}(b_{1})\big) \Big\} .
\end{split}
\label{eq:dtau}
\end{equation}
This expression is numerically stable as long as $b_1$ is not too close to $b_0$, but will suffer from cancellation errors otherwise.
Therefore, for $\left|b_{1}-b_{0}\right| < 10^{-3}$, we use the first two terms of the Taylor expansion of (\ref{eq:dtau}) in $b_1$ around $b_0$, instead of equation (\ref{eq:dtau}),
\begin{equation}
\Delta \tau \ \approx \ \Delta z \, e^{-b_{0}^{2}} \left(\frac{1}{2}\left(a_{0} + a_{1}\right) \ - \ \frac{1}{3} \, \left( a_{0} + 2 a_{1} \right) b_{0} \left(b_{1}-b_{0} \right) \right) .
\end{equation}
This implementation of the line optical depth can be found in the \textsc{lines} class in \textsc{pomme}\footnote{See also \url{https://github.com/Magritte-code/pomme/blob/main/src/pomme/lines.py}.}.
It turns out that an implementation with two masks (one for the case where $\left|b_{1}-b_{0}\right| < 10^{-3}$ and one for its complement) is computationally more expensive than performing the calculations for both cases, and only in the end merging the results. This, however, means that at some point $\Delta \tau$ will be undefined due to division by zero (as $b_{1}-b_{0}$ approaches zero).
This causes no problem for the forward model (since these values will eventually be overwritten), but it will cause gradients to diverge in \textsc{PyTorch}, which is a problem when using these to solve the inverse problem.
To avoid this, we add a small number ($10^{-30}$) to the denominator in (\ref{eq:dtau}).

\subsection{Radiative transfer}
\label{app:radiative_transfer}
Next, we consider the implementation of the formal solution of the radiative transfer problem, as given in equation (\ref{eq:formalsolution}).
This is done in a way similar to the optical depth by solving the integral analytically after a local assumption of linearity.

Consider again a line-of-sight segment between two consecutive elements, indexed as 0 and 1, parametrised by $\lambda \in [0, 1]$.
The accumulated intensity in this segment can then be written as,
\begin{equation}
\Delta I \ = \ \Delta\tau \int_{0}^{1} \text{d}\lambda \ S(\lambda) \, e^{-\tau(\lambda)},
\end{equation}
where the source function is defined as, $S(\lambda) \equiv \eta(\lambda) / \chi(\lambda)$.
Using linear interpolation for the source function, $S$, and the optical depth, $\tau$,
\begin{align}
S   (\lambda) \ &= \ (1-\lambda)    S_{0} \ + \ \lambda    S_{1}, \\
\tau(\lambda) \ &= \ (1-\lambda) \tau_{0} \ + \ \lambda \tau_{1},
\end{align}
with discretised values $S_{0}$ and $\tau_{0}$ at element 0, and with discretised values $S_{1}$ and $\tau_{1}$ at element 1,
the formal solution yields,
\begin{equation}
\Delta I \ = \ \frac{1}{\Delta\tau} \Big( S_{0} \, e^{-\tau_{0}} \left( e^{-\Delta\tau} - (1 - \Delta\tau) \right) \ + \
S_{1} \, e^{-\tau_{1}} \left( e^{+\Delta\tau} - (1 + \Delta\tau) \right) \Big),
\end{equation}
where $\Delta\tau \equiv \tau_{1} - \tau_{0}$. This expression is numerically stable as long as $\Delta \tau$ is not too small, but will suffer from cancellation errors otherwise.
Therefore, for $|\Delta \tau| < 10^{-2}$, we use the first three terms in the Taylor expansion,
\begin{equation}
\frac{1}{\Delta\tau} \left(e^{\mp\Delta\tau} - (1 \mp \Delta\tau) \right) \ \approx \ \frac{1}{2}\Delta\tau \ \mp \ \frac{1}{6}\Delta\tau^{2} \ + \ \frac{1}{24}\Delta\tau^{3},
\end{equation}
where we recognise the expansion of the exponential minus the first two terms.
This implementation of the intensity increment can be found in the \textsc{forward} class in \textsc{pomme}\footnote{See also \url{https://github.com/Magritte-code/pomme/blob/main/src/pomme/forward.py}.}.
For the same reason as with the optical depth (see Appendix \ref{app:optical_depth}), we compute both cases for all values, to minimise the use of masks, and we add a small number ($10^{-30}$) to the denominator to avoid division by zero and the resulting undefined numbers when computing gradients.

\subsection{Spherically symmetric models}
Within \textsc{pomme}, we provide some specific functionality for modelling spherically symmetric models.
Since spherically symmetric models are essentially one dimensional, they can be defined in terms of a set of 1D tensors.
However, since radiation not necessarily propagates along the radial direction but can also propagate at an angle, the radiative transfer problem, even under the assumption of spherically symmetry, is still a 2D problem.
To solve the radiative transfer problem along different rays we use a new \textsc{Python} implementation of the 1D ray-tracer for spherically symmetric models as implemented in \textsc{Magritte}\footnote{The source code can be found at: \url{https://github.com/Magritte-code/Magritte}.} \citep[][{\color{black}Ceulemans et al.\ \textit{subm.}}]{de_ceuster_magritte_2019, de_ceuster_magritte_2020, de_ceuster_3d_2022}. The ray-tracer provides the indices of relevant model data in the 1D tensors and the corresponding distance increments.
These are then fed into the line integral solvers (with variable distance increments, $\Delta x$) and solved as described in Appendices \ref{app:optical_depth} and \ref{app:radiative_transfer}.
Finally, the resulting intensities are integrated over all impact factors to produce the spectral energy distribution\footnote{See also \url{https://github.com/Magritte-code/pomme/blob/main/src/pomme/model.py}.}.

\section{Additional tables \& figures}
\label{app:figures}
In this appendix, we provide some additional tables and figures.

\begin{deluxetable}{l | l | c | c}[h!]
\tablecaption{Summary of the reconstructions presented in this paper.}
\label{tab:sum}
\tablehead{ Sect. & Model type & Free parameters & Prior}
\startdata
\ref{subsubsec:Mdot} & 1D analytic (sph.\ sym.) & $\dot{M}$, $v_{\star}$, $v_{\infty}$, $\beta$, $T_{\star}$, $\epsilon$ & none \rule{0pt}{13pt} \\
\ref{subsubsec:only_CO} & 1D analytic (sph.\ sym.) & $n_{\text{CO}}(r)$ & smoothness (\ref{eq:reg_smooth}) + continuity (\ref{eq:1D_cont}) \rule{0pt}{13pt} \\
\ref{subsubsec:all} & 1D analytic (sph.\ sym.)  & $n_{\text{CO}}(r)$, $T(r)$, $v(r)$ & smoothness (\ref{eq:reg_smooth}) + continuity (\ref{eq:1D_cont}) \rule{0pt}{13pt} \\
\ref{subsec:phantom} & 3D numeric & $\rho(\boldsymbol{x})$, $T(\boldsymbol{x})$, $v_{r}(\boldsymbol{x})$ & smoothness (\ref{eq:reg_smooth}) + continuity (\ref{eq:3D_cont})  \rule{0pt}{13pt} \\
\ref{subsubsec:rec_NaCl} & 3D numeric & $n_{\text{NaCl}}(\boldsymbol{x})$  & smoothness (\ref{eq:reg_smooth})  \rule{0pt}{13pt}
\enddata
\end{deluxetable}

\begin{figure}[!ph]
    \centering
    \includegraphics[width=\linewidth]{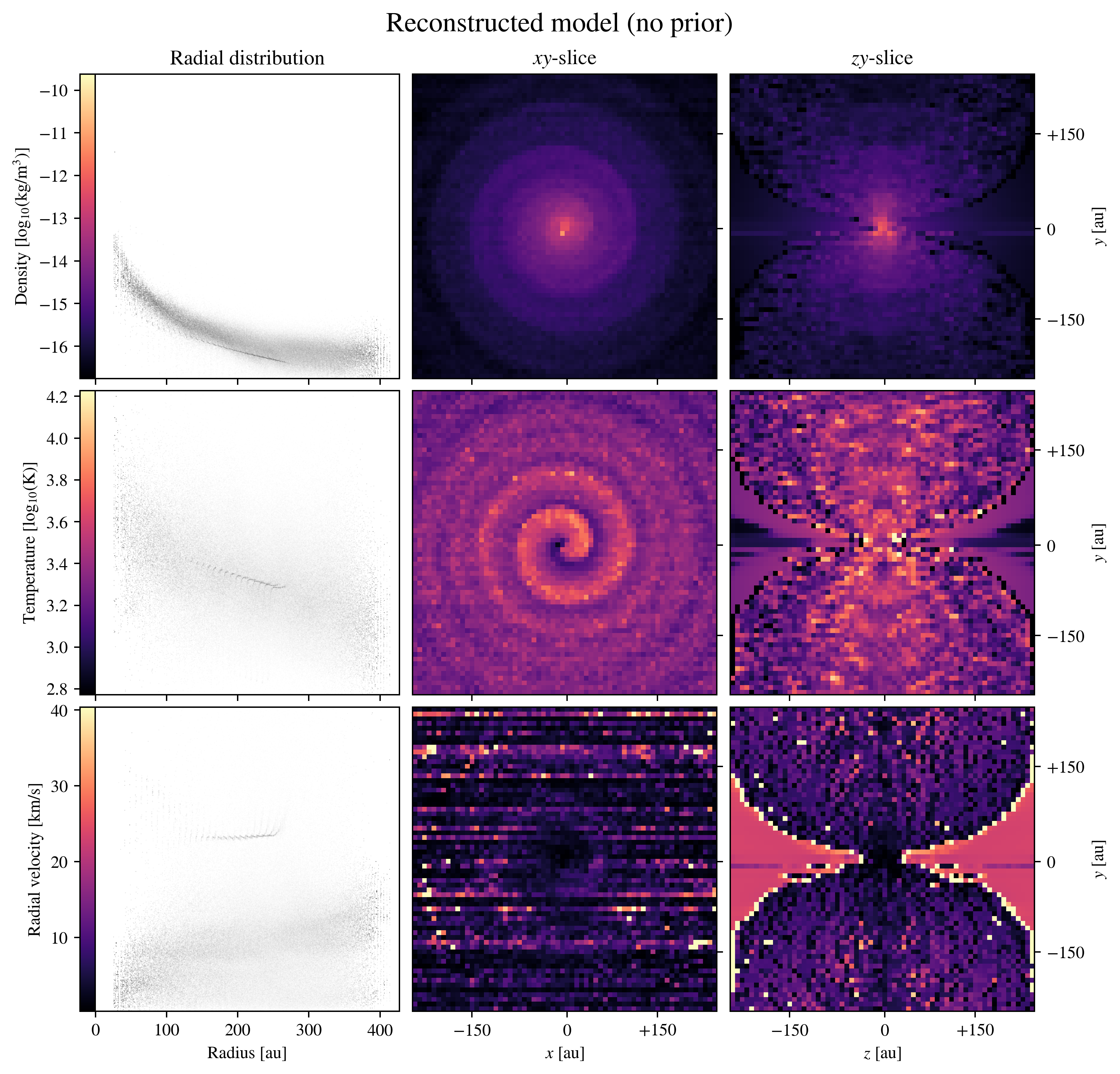}
    \caption{
        Radial distribution and two slices of the reconstruction (without any prior) of the density, temperature, and radial velocity distributions of the companion-perturbed AGB wind hydrodynamics model. The radial distributions are obtained with 2D histograms with $428^{2}$ bins, evenly spaced within the parameter range of the original model. The colour bars also represent the vertical axis of the radial distributions.
    }
    \label{fig:phantom_rec_nono}
\end{figure}

\begin{figure}[!ph]
    \centering
    \includegraphics[width=\linewidth]{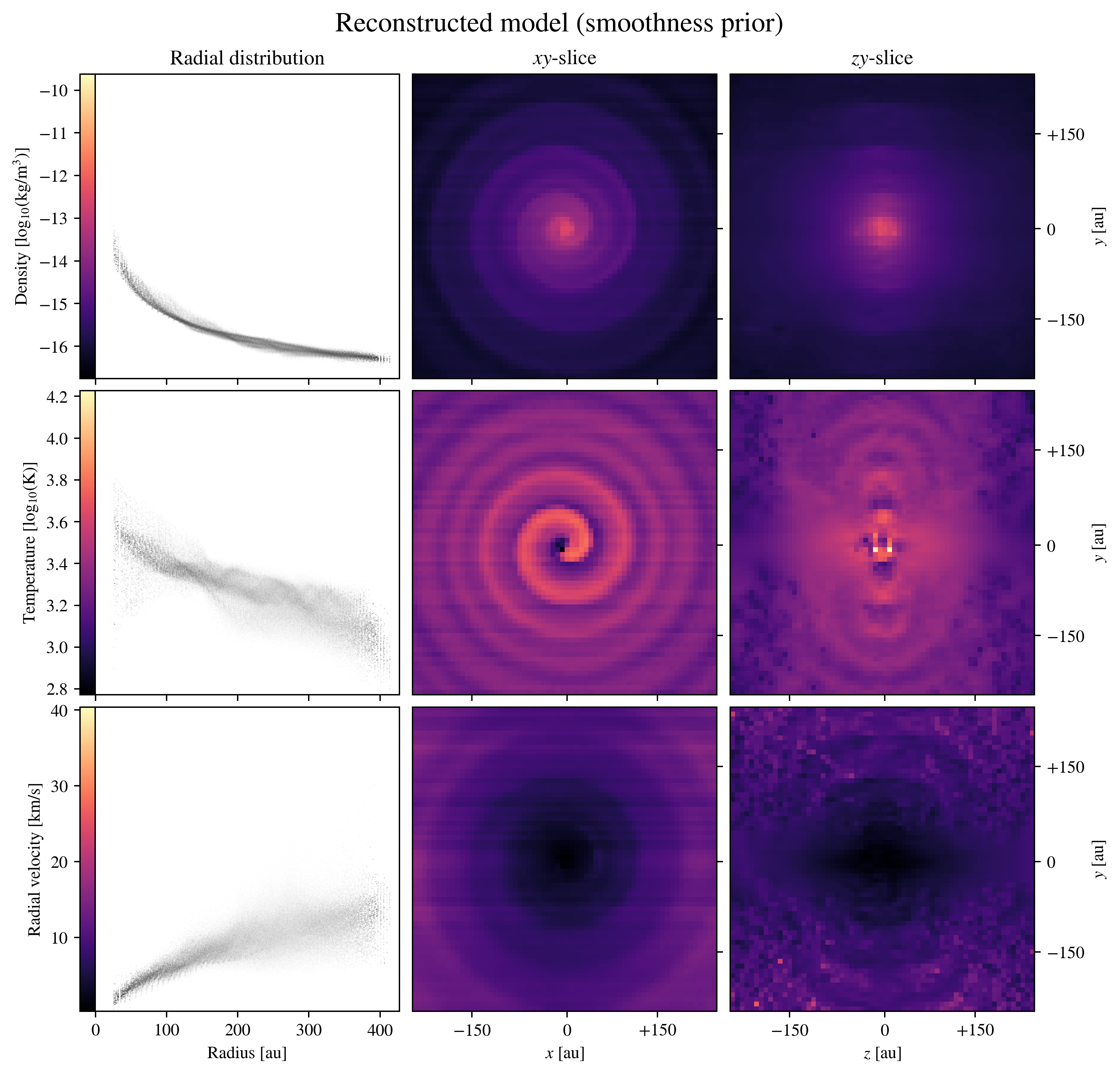}
    \caption{
        Radial distribution and two slices of the reconstruction (only using a smoothness prior) of the density, temperature, and radial velocity distributions of the companion-perturbed AGB wind hydrodynamics model. The radial distributions are obtained with 2D histograms with $428^{2}$ bins, evenly spaced within the parameter range of the original model. The colour bars also represent the vertical axis of the radial distributions.
    }
    \label{fig:phantom_rec_noco}
\end{figure}

\begin{figure}[!ph]
    \centering
    \includegraphics[width=0.93\linewidth]{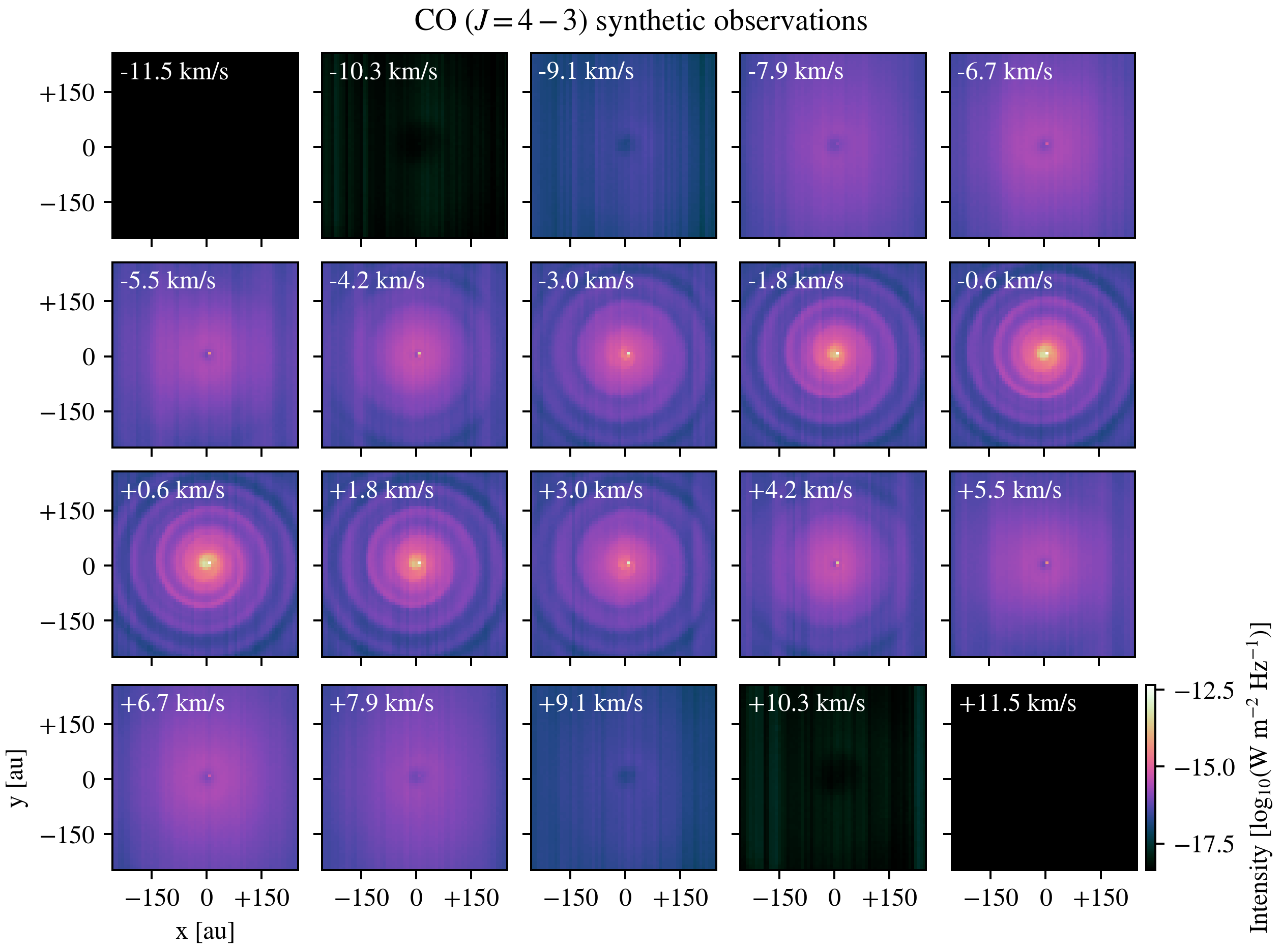}
    \includegraphics[width=0.93\linewidth]{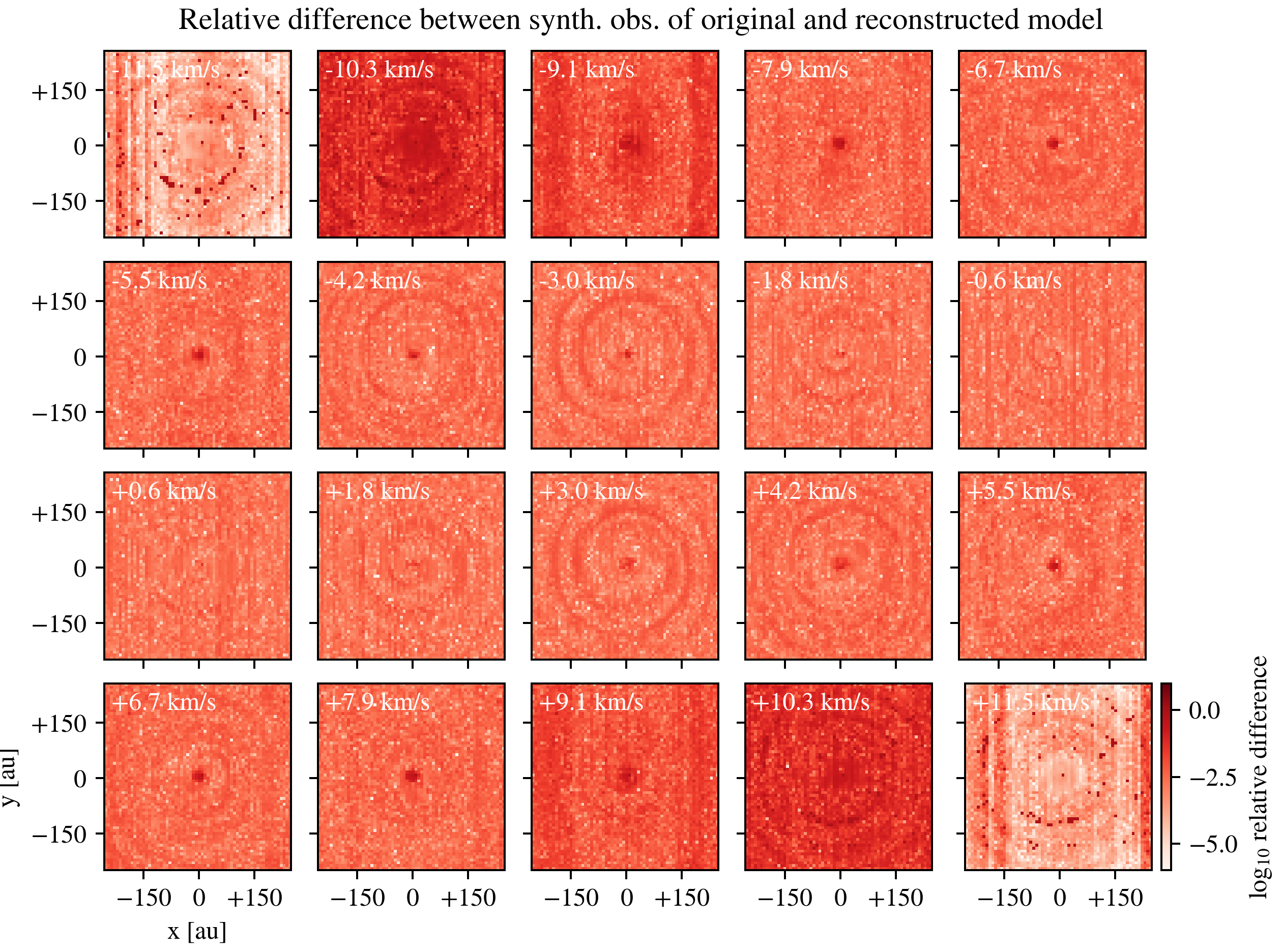}
    \caption{
        \textit{(Top.)} Synthetic observations of the CO($J=4-3$) line of the companion-perturbed AGB wind model that we aim to reconstruct in Section \ref{subsec:phantom}. \textit{(Bottom.)} Relative differences between the synthetic observations of the original and reconstructed model.
        Note that the figure only shows 20 of the total 100 frequency bins.
    }
    \label{fig:phantom_obs_CO}
\end{figure}

\begin{figure}[!ph]
    \centering
    \includegraphics[width=0.93\linewidth]{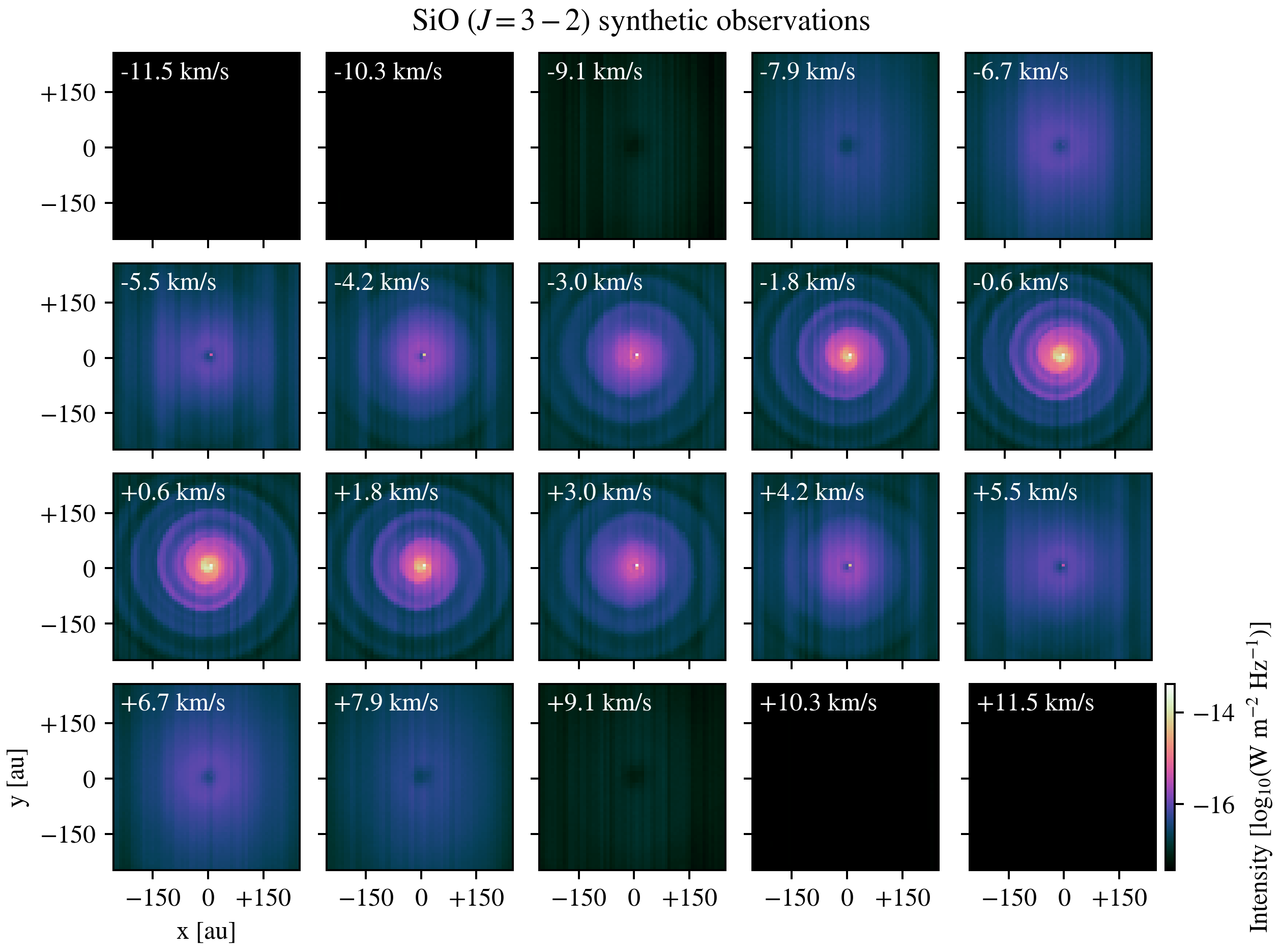}
    \includegraphics[width=0.93\linewidth]{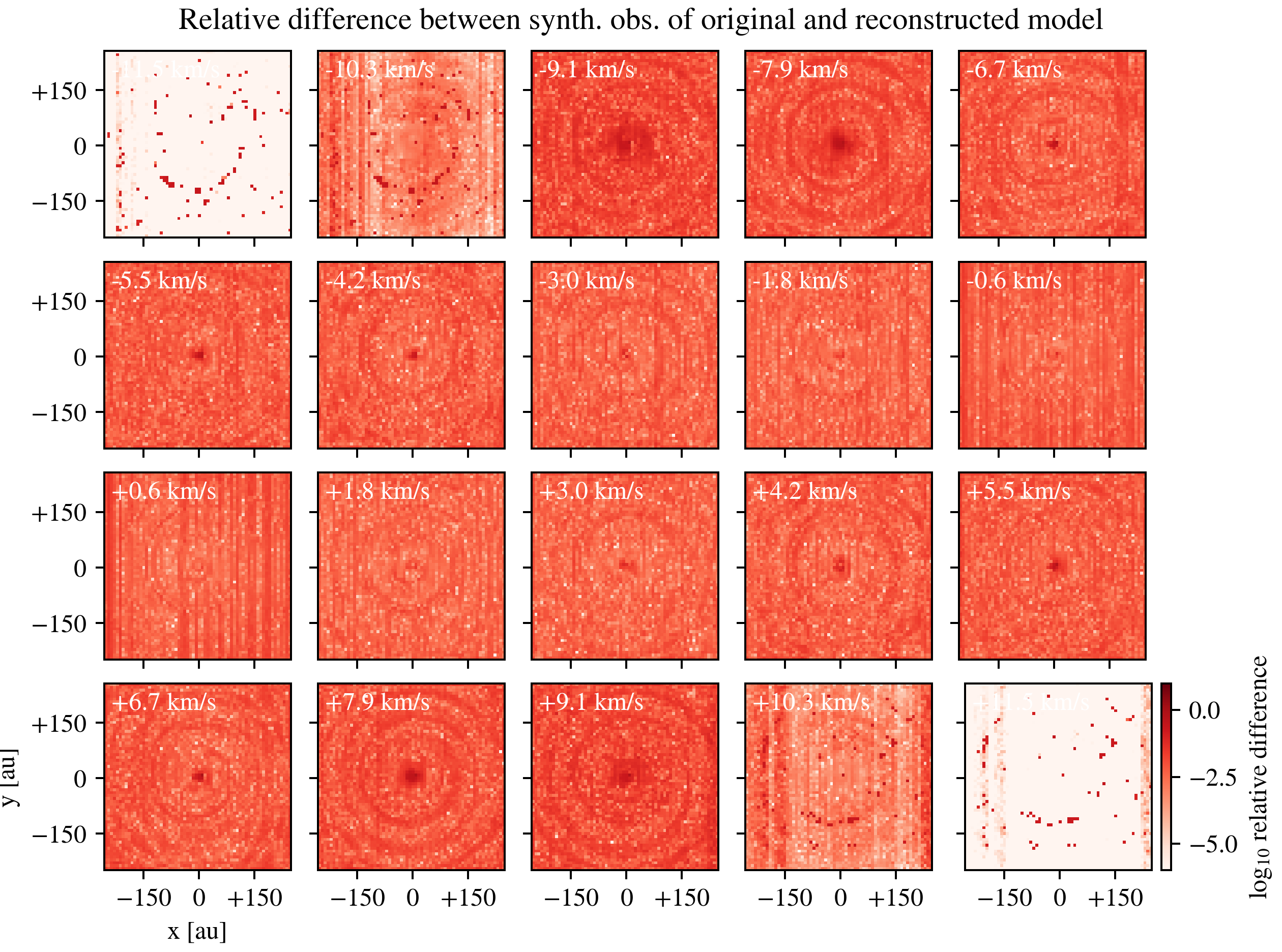}
    \caption{
        \textit{(Top.)} Synthetic observations of the SiO($J=3-2$) line of the companion-perturbed AGB wind model that we aim to reconstruct in Section \ref{subsec:phantom}.
        \textit{(Bottom.)} Relative differences between the synthetic observations of the original and reconstructed model.
        Note that the figure only shows 20 of the total 100 frequency bins.
    }
    \label{fig:phantom_obs_SiO}
\end{figure}

\begin{figure}[!ph]
    \centering
    \includegraphics[width=0.93\linewidth]{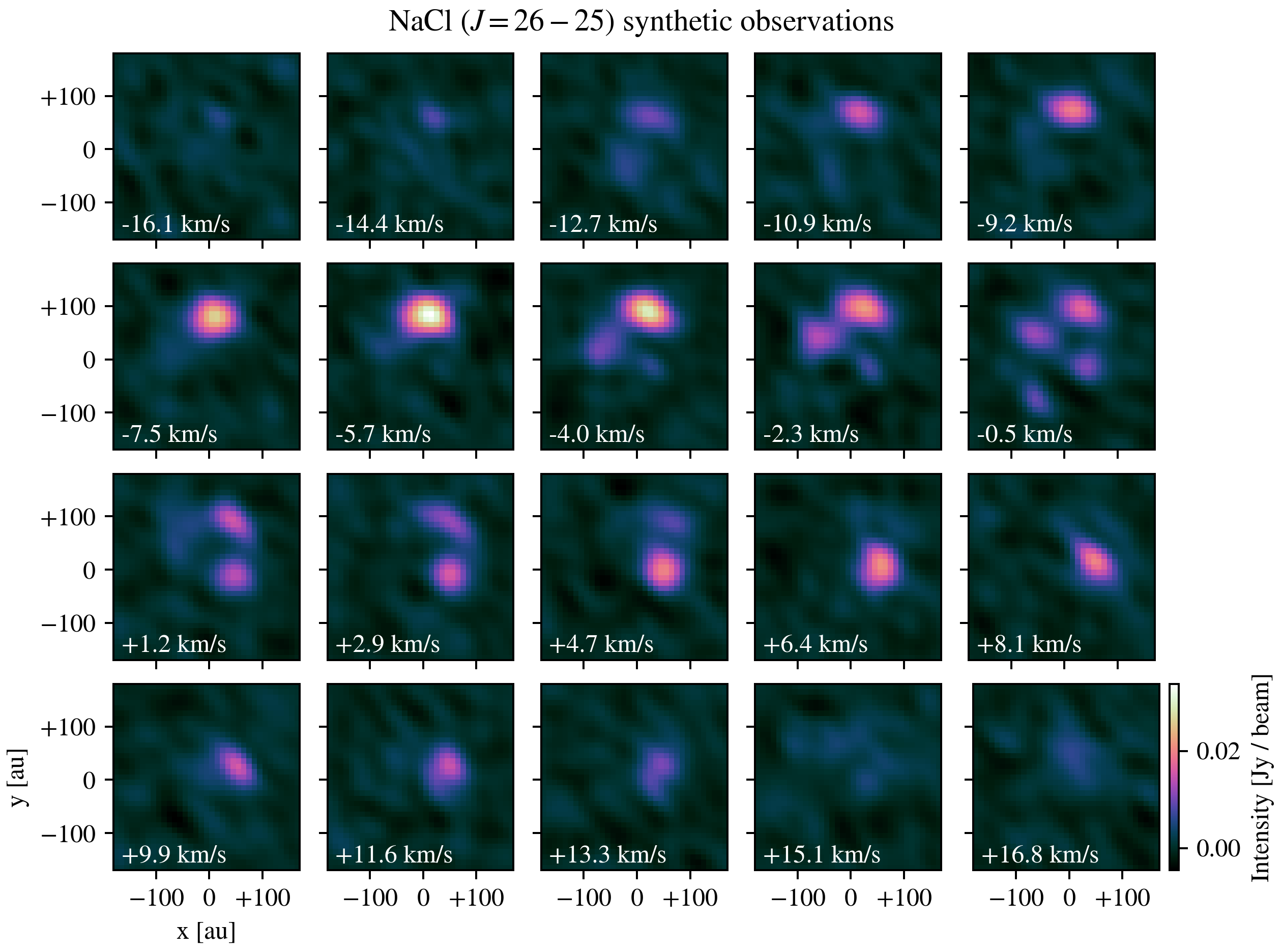}
    \includegraphics[width=0.93\linewidth]{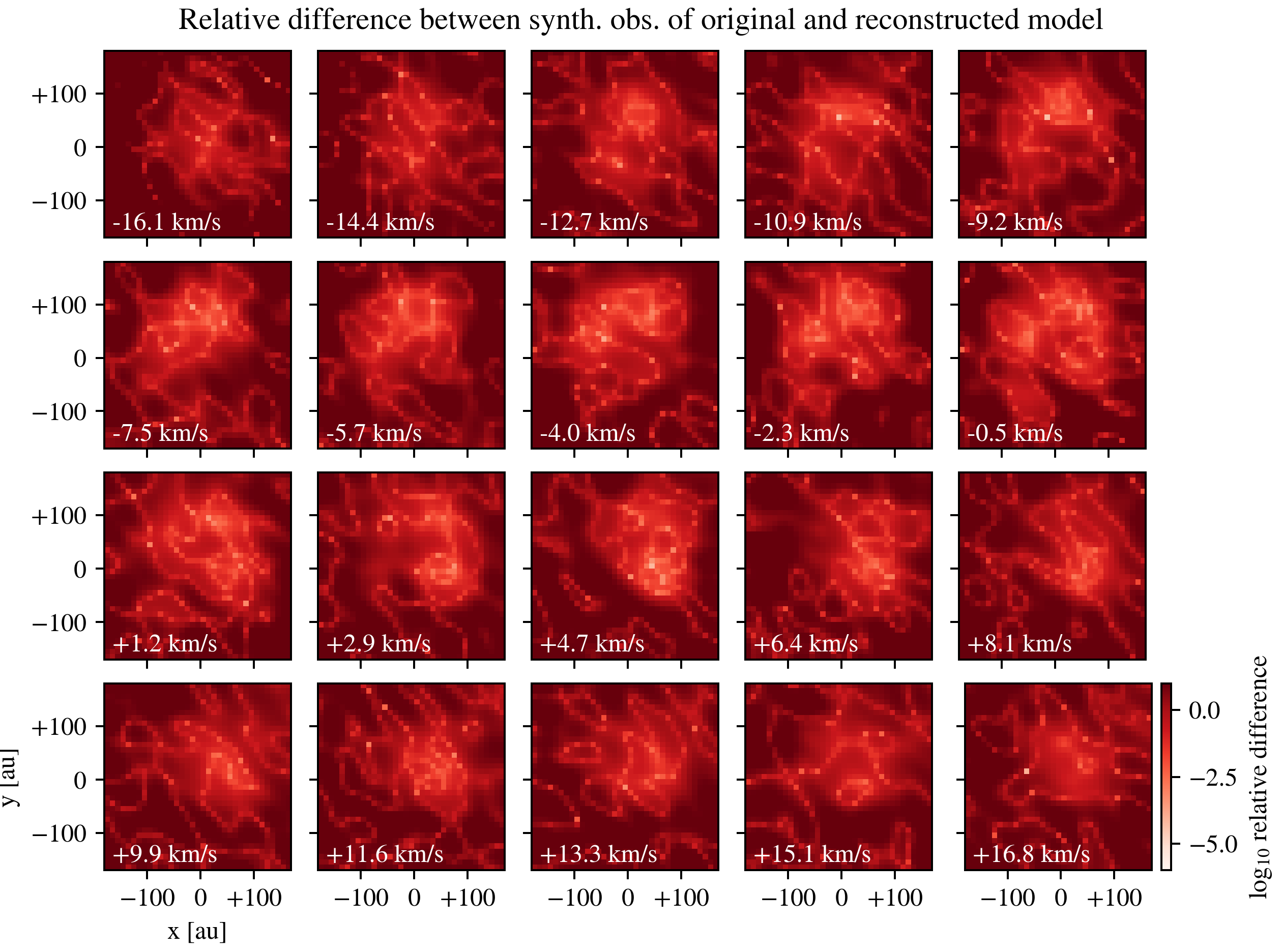}
    \caption{
        \textit{(Top.)} Observations of the NaCl($J=26-25$) line around the AGB star IK Tau that we aim to reconstruct in Section \ref{sec:applications}.
        \textit{(Bottom.)} Relative differences between the true observations and the synthetic observations of the reconstructed model.
        Note that the figure only shows 20 of the total 24 frequency bins.
    }
    \label{fig:IKTau_obs}
\end{figure}

\newpage
\bibliography{references}{}

\begin{thebibliography}{}
\expandafter\ifx\csname natexlab\endcsname\relax\def\natexlab#1{#1}\fi
\providecommand{\url}[1]{\href{#1}{#1}}
\providecommand{\dodoi}[1]{doi:~\href{http://doi.org/#1}{\nolinkurl{#1}}}
\providecommand{\doeprint}[1]{\href{http://ascl.net/#1}{\nolinkurl{http://ascl.net/#1}}}
\providecommand{\doarXiv}[1]{\href{https://arxiv.org/abs/#1}{\nolinkurl{https://arxiv.org/abs/#1}}}

\bibitem[{Abellán {et~al.}(2017)Abellán, Indebetouw, Marcaide, Gabler,
  Fransson, Spyromilio, Burrows, Chevalier, Cigan, Gaensler, Gomez, Janka,
  Kirshner, Larsson, Lundqvist, Matsuura, McCray, Ng, Park, Roche,
  Staveley-Smith, van Loon, Wheeler, \& Woosley}]{abellan_very_2017}
Abellán, F.~J., Indebetouw, R., Marcaide, J.~M., {et~al.} 2017, The
  Astrophysical Journal, 842, L24, \dodoi{10.3847/2041-8213/aa784c}

\bibitem[{Afkham {et~al.}(2021)Afkham, Chung, \& Chung}]{afkham_learning_2021}
Afkham, B.~M., Chung, J., \& Chung, M. 2021, Inverse Problems, 37, 105017,
  \dodoi{10.1088/1361-6420/ac245d}

\bibitem[{Andrews {et~al.}(2018)Andrews, Huang, Pérez, Isella, Dullemond,
  Kurtovic, Guzmán, Carpenter, Wilner, Zhang, Zhu, Birnstiel, Bai, Benisty,
  Hughes, Öberg, \& Ricci}]{andrews_disk_2018}
Andrews, S.~M., Huang, J., Pérez, L.~M., {et~al.} 2018, The Astrophysical
  Journal, 869, L41, \dodoi{10.3847/2041-8213/aaf741}

\bibitem[{Asensio~Ramos \& de~la
  Cruz~Rodríguez(2015)}]{asensio_ramos_sparse_2015}
Asensio~Ramos, A., \& de~la Cruz~Rodríguez, J. 2015, Astronomy \&
  Astrophysics, 577, A140, \dodoi{10.1051/0004-6361/201425508}

\bibitem[{Asensio~Ramos {et~al.}(2022)Asensio~Ramos, Díaz~Baso, \&
  Kochukhov}]{asensio_ramos_approximate_2022}
Asensio~Ramos, A., Díaz~Baso, C.~J., \& Kochukhov, O. 2022, Astronomy \&
  Astrophysics, 658, A162, \dodoi{10.1051/0004-6361/202142027}

\bibitem[{Asensio~Ramos {et~al.}(2007)Asensio~Ramos, González~Martínez, \&
  Rubiño-Martín}]{asensio_ramos_bayesian_2007}
Asensio~Ramos, A., González~Martínez, M.~J., \& Rubiño-Martín, J.~A. 2007,
  Astronomy \& Astrophysics, 476, 959, \dodoi{10.1051/0004-6361:20078107}

\bibitem[{Balakrishnan {et~al.}(2019)Balakrishnan, Dalca, Zhao, Guttag, Durand,
  \& Freeman}]{balakrishnan_visual_2019}
Balakrishnan, G., Dalca, A.~V., Zhao, A., {et~al.} 2019, 171--180.
\newblock
  \url{https://openaccess.thecvf.com/content_ICCV_2019/html/Balakrishnan_Visual_Deprojection_Probabilistic_Recovery_of_Collapsed_Dimensions_ICCV_2019_paper.html}

\bibitem[{Bertero {et~al.}(2021)Bertero, Boccacci, \&
  Mol}]{bertero_introduction_2021}
Bertero, M., Boccacci, P., \& Mol, C.~D. 2021, Introduction to {Inverse}
  {Problems} in {Imaging}, 2nd edn. (Boca Raton: CRC Press),
  \dodoi{10.1201/9781003032755}

\bibitem[{Bitsch {et~al.}(2015)Bitsch, Johansen, Lambrechts, \&
  Morbidelli}]{bitsch_structure_2015}
Bitsch, B., Johansen, A., Lambrechts, M., \& Morbidelli, A. 2015, Astronomy \&
  Astrophysics, 575, A28, \dodoi{10.1051/0004-6361/201424964}

\bibitem[{Chen {et~al.}(2016)Chen, Nordhaus, Frank, Blackman, \&
  Balick}]{chen_three-dimensional_2016}
Chen, Z., Nordhaus, J., Frank, A., Blackman, E.~G., \& Balick, B. 2016, Monthly
  Notices of the Royal Astronomical Society, 460, 4182,
  \dodoi{10.1093/mnras/stw1305}

\bibitem[{Coenegrachts {et~al.}(2023)Coenegrachts, Danilovich, De~Ceuster, \&
  Decin}]{coenegrachts_unusual_2023}
Coenegrachts, A., Danilovich, T., De~Ceuster, F., \& Decin, L. 2023, The
  unusual {3D} distribution of {NaCl} around the {AGB} star {IK} {Tau},
  \dodoi{10.48550/arXiv.2302.06221}

\bibitem[{Danilovich {et~al.}(2024)Danilovich, Malfait, Van~de Sande,
  Montargès, Kervella, De~Ceuster, Coenegrachts, Millar, Richards, Decin,
  Gottlieb, Pinte, De~Beck, Price, Wong, Bolte, Menten, Baudry, de~Koter,
  Etoka, Gobrecht, Gray, Herpin, Jeste, Lagadec, Maes, McDonald, Marinho,
  Müller, Pimpanuwat, Plane, Sahai, Wallström, Yates, \&
  Zijlstra}]{danilovich_chemical_2024}
Danilovich, T., Malfait, J., Van~de Sande, M., {et~al.} 2024, Nature Astronomy,
  \dodoi{10.1038/s41550-023-02154-y}

\bibitem[{De~Ceuster {et~al.}(2023)De~Ceuster, Ceulemans, Cockayne, Decin, \&
  Yates}]{de_ceuster_radiative_2023}
De~Ceuster, F., Ceulemans, T., Cockayne, J., Decin, L., \& Yates, J. 2023,
  Monthly Notices of the Royal Astronomical Society, 518, 5536,
  \dodoi{10.1093/mnras/stac3461}

\bibitem[{De~Ceuster {et~al.}(2019)De~Ceuster, Homan, Yates, Decin, Boyle, \&
  Hetherington}]{de_ceuster_magritte_2019}
De~Ceuster, F., Homan, W., Yates, J., {et~al.} 2019, Monthly Notices of the
  Royal Astronomical Society, 492, 1812, \dodoi{10.1093/mnras/stz3557}

\bibitem[{De~Ceuster {et~al.}(2020)De~Ceuster, Bolte, Homan, Maes, Malfait,
  Decin, Yates, Boyle, \& Hetherington}]{de_ceuster_magritte_2020}
De~Ceuster, F., Bolte, J., Homan, W., {et~al.} 2020, Monthly Notices of the
  Royal Astronomical Society, 499, 5194, \dodoi{10.1093/mnras/staa3199}

\bibitem[{De~Ceuster {et~al.}(2022)De~Ceuster, Ceulemans, Srivastava, Homan,
  Bolte, Yates, Decin, Boyle, \& Hetherington}]{de_ceuster_3d_2022}
De~Ceuster, F., Ceulemans, T., Srivastava, A., {et~al.} 2022, Journal of Open
  Source Software, 7, 3905, \dodoi{10.21105/joss.03905}

\bibitem[{de~Mijolla {et~al.}(2019)de~Mijolla, Viti, Holdship, Manolopoulou, \&
  Yates}]{de_mijolla_incorporating_2019}
de~Mijolla, D., Viti, S., Holdship, J., Manolopoulou, I., \& Yates, J. 2019,
  Astronomy \& Astrophysics, 630, A117, \dodoi{10.1051/0004-6361/201935973}

\bibitem[{Decin {et~al.}(2020)Decin, Montargès, Richards, Gottlieb, Homan,
  McDonald, El~Mellah, Danilovich, Wallström, Zijlstra, Baudry, Bolte, Cannon,
  De~Beck, De~Ceuster, de~Koter, De~Ridder, Etoka, Gobrecht, Gray, Herpin,
  Jeste, Lagadec, Kervella, Khouri, Menten, Millar, Müller, Plane, Sahai,
  Sana, Van~de Sande, Waters, Wong, \& Yates}]{decin_substellar_2020}
Decin, L., Montargès, M., Richards, A. M.~S., {et~al.} 2020, Science, 369,
  1497, \dodoi{10.1126/science.abb1229}

\bibitem[{del Toro~Iniesta \&
  Ruiz~Cobo(2016)}]{del_toro_iniesta_inversion_2016}
del Toro~Iniesta, J.~C., \& Ruiz~Cobo, B. 2016, Living Reviews in Solar
  Physics, 13, 4, \dodoi{10.1007/s41116-016-0005-2}

\bibitem[{de la Cruz~Rodríguez \& van
  Noort(2017)}]{dela_cruz_rodriguez_radiative_2017}
de la Cruz~Rodríguez, J., \& van Noort, M. 2017, Space Science Reviews, 210,
  109, \dodoi{10.1007/s11214-016-0294-8}

\bibitem[{Díaz~Baso {et~al.}(2022)Díaz~Baso, Asensio~Ramos, \& de~la
  Cruz~Rodrígez}]{diaz_baso_bayesian_2022}
Díaz~Baso, C.~J., Asensio~Ramos, A., \& de~la Cruz~Rodrígez, J. 2022,
  Astronomy \& Astrophysics, 659, A165, \dodoi{10.1051/0004-6361/202142018}

\bibitem[{Esseldeurs {et~al.}(2023)Esseldeurs, Siess, De~Ceuster, Homan,
  Malfait, Maes, Konings, Ceulemans, \& Decin}]{esseldeurs_3d_2023}
Esseldeurs, M., Siess, L., De~Ceuster, F., {et~al.} 2023, Astronomy and
  Astrophysics, 674, A122, \dodoi{10.1051/0004-6361/202346282}

\bibitem[{Guélin {et~al.}(2018)Guélin, Patel, Bremer, Cernicharo,
  Castro-Carrizo, Pety, Fonfría, Agúndez, Santander-García, Quintana-Lacaci,
  Velilla~Prieto, Blundell, \& Thaddeus}]{guelin_irc_2018}
Guélin, M., Patel, N.~A., Bremer, M., {et~al.} 2018, Astronomy and
  Astrophysics, 610, A4, \dodoi{10.1051/0004-6361/201731619}

\bibitem[{Haber \& Tenorio(2003)}]{haber_learning_2003}
Haber, E., \& Tenorio, L. 2003, Inverse Problems, 19, 611,
  \dodoi{10.1088/0266-5611/19/3/309}

\bibitem[{Homan {et~al.}(2018{\natexlab{a}})Homan, Danilovich, Decin, de~Koter,
  Nuth, \& Van~de Sande}]{homan_alma_2018}
Homan, W., Danilovich, T., Decin, L., {et~al.} 2018{\natexlab{a}}, Astronomy
  and Astrophysics, 614, A113, \dodoi{10.1051/0004-6361/201732246}

\bibitem[{Homan {et~al.}(2018{\natexlab{b}})Homan, Richards, Decin, de~Koter,
  \& Kervella}]{homan_unusual_2018}
Homan, W., Richards, A., Decin, L., de~Koter, A., \& Kervella, P.
  2018{\natexlab{b}}, Astronomy and Astrophysics, 616, A34,
  \dodoi{10.1051/0004-6361/201832834}

\bibitem[{Hubert {et~al.}(2016)Hubert, Opitom, Hutsemékers, Jehin, Munhoven,
  Manfroid, Bisikalo, \& Shematovich}]{hubert_inversion_2016}
Hubert, B., Opitom, C., Hutsemékers, D., {et~al.} 2016, Icarus, 277, 237,
  \dodoi{10.1016/j.icarus.2016.04.044}

\bibitem[{Högbom(1974)}]{hogbom_aperture_1974}
Högbom, J.~A. 1974, Astronomy and Astrophysics Supplement Series, 15, 417.
\newblock \url{https://ui.adsabs.harvard.edu/abs/1974A&AS...15..417H}

\bibitem[{Kaastra(1989)}]{kaastra_deprojection_1989}
Kaastra, J.~S. 1989, Astronomy and Astrophysics, 224, 338.
\newblock \url{https://ui.adsabs.harvard.edu/abs/1989A&A...224..338K}

\bibitem[{Kervella {et~al.}(2016)Kervella, Homan, Richards, Decin, McDonald,
  Montargès, \& Ohnaka}]{kervella_alma_2016}
Kervella, P., Homan, W., Richards, A. M.~S., {et~al.} 2016, Astronomy and
  Astrophysics, 596, A92, \dodoi{10.1051/0004-6361/201629877}

\bibitem[{Kingma \& Ba(2015)}]{kingma_adam_2015}
Kingma, D.~P., \& Ba, J. 2015, Adam: {A} {Method} for {Stochastic}
  {Optimization},  arXiv, \dodoi{10.48550/arXiv.1412.6980}

\bibitem[{Ksoll {et~al.}(2023)Ksoll, Reissl, Klessen, Stephens, Smith, Soler,
  Traficante, Testi, Hennebelle, \& Molinari}]{ksoll_deep_2023}
Ksoll, V.~F., Reissl, S., Klessen, R.~S., {et~al.} 2023, A deep learning
  approach for the {3D} reconstruction of dust density and temperature in
  star-forming regions,  arXiv.
\newblock \url{http://arxiv.org/abs/2308.09657}

\bibitem[{Lee {et~al.}(2022)Lee, Kim, \& Lee}]{lee_formation_2022}
Lee, Y.-M., Kim, H., \& Lee, H.-W. 2022, The Astrophysical Journal, 931, 142,
  \dodoi{10.3847/1538-4357/ac67d6}

\bibitem[{Lucy(1974)}]{lucy_iterative_1974}
Lucy, L.~B. 1974, The Astronomical Journal, 79, 745, \dodoi{10.1086/111605}

\bibitem[{Maercker {et~al.}(2012)Maercker, Mohamed, Vlemmings, Ramstedt,
  Groenewegen, Humphreys, Kerschbaum, Lindqvist, Olofsson, Paladini,
  Wittkowski, de~Gregorio-Monsalvo, \& Nyman}]{maercker_unexpectedly_2012}
Maercker, M., Mohamed, S., Vlemmings, W. H.~T., {et~al.} 2012, Nature, 490,
  232, \dodoi{10.1038/nature11511}

\bibitem[{Maes {et~al.}(2021)Maes, Homan, Malfait, Siess, Bolte, De~Ceuster, \&
  Decin}]{maes_sph_2021}
Maes, S., Homan, W., Malfait, J., {et~al.} 2021, Astronomy and Astrophysics,
  653, A25, \dodoi{10.1051/0004-6361/202140823}

\bibitem[{Malfait {et~al.}(2021)Malfait, Homan, Maes, Bolte, Siess, De~Ceuster,
  \& Decin}]{malfait_sph_2021}
Malfait, J., Homan, W., Maes, S., {et~al.} 2021, Astronomy and Astrophysics,
  652, A51, \dodoi{10.1051/0004-6361/202141161}

\bibitem[{Mauron \& Huggins(2006)}]{mauron_imaging_2006}
Mauron, N., \& Huggins, P.~J. 2006, Astronomy and Astrophysics, 452, 257,
  \dodoi{10.1051/0004-6361:20054739}

\bibitem[{Milisavljevic \& Fesen(2015)}]{milisavljevic_bubble-like_2015}
Milisavljevic, D., \& Fesen, R.~A. 2015, Science, 347, 526,
  \dodoi{10.1126/science.1261949}

\bibitem[{Montargès {et~al.}(2019)Montargès, Homan, Keller, Clementel,
  Shetye, Decin, Harper, Royer, Winters, Le~Bertre, \&
  Richards}]{montarges_noema_2019}
Montargès, M., Homan, W., Keller, D., {et~al.} 2019, Monthly Notices of the
  Royal Astronomical Society, 485, 2417, \dodoi{10.1093/mnras/stz397}

\bibitem[{Palmer(1994)}]{palmer_deprojection_1994}
Palmer, P.~L. 1994, Monthly Notices of the Royal Astronomical Society, 266,
  697, \dodoi{10.1093/mnras/266.3.697}

\bibitem[{Paszke {et~al.}(2017)Paszke, Gross, Chintala, Chanan, Yang, DeVito,
  Lin, Desmaison, Antiga, \& Lerer}]{paszke_automatic_2017}
Paszke, A., Gross, S., Chintala, S., {et~al.} 2017, NIPS 2017 Autodiff
  Workshop.
\newblock \url{https://openreview.net/forum?id=BJJsrmfCZ}

\bibitem[{Paszke {et~al.}(2019)Paszke, Gross, Massa, Lerer, Bradbury, Chanan,
  Killeen, Lin, Gimelshein, Antiga, Desmaison, Kopf, Yang, DeVito, Raison,
  Tejani, Chilamkurthy, Steiner, Fang, Bai, \& Chintala}]{paszke_pytorch_2019}
Paszke, A., Gross, S., Massa, F., {et~al.} 2019, in Advances in {Neural}
  {Information} {Processing} {Systems} 32 ({NeurIPS} 2019), Vol.~32 (Curran
  Associates, Inc.).
\newblock
  \url{https://papers.nips.cc/paper_files/paper/2019/hash/bdbca288fee7f92f2bfa9f7012727740-Abstract.html}

\bibitem[{Ramstedt {et~al.}(2014)Ramstedt, Mohamed, Vlemmings, Maercker,
  Montez, Baudry, De~Beck, Lindqvist, Olofsson, Humphreys, Jorissen,
  Kerschbaum, Mayer, Wittkowski, Cox, Lagadec, Leal-Ferreira, Paladini,
  Pérez-Sánchez, \& Sacuto}]{ramstedt_wonderful_2014}
Ramstedt, S., Mohamed, S., Vlemmings, W. H.~T., {et~al.} 2014, Astronomy and
  Astrophysics, 570, L14, \dodoi{10.1051/0004-6361/201425029}

\bibitem[{Rybicki(1987)}]{rybicki_deprojection_1987}
Rybicki, G.~B. 1987, 127, 397, \dodoi{10.1007/978-94-009-3971-4_41}

\bibitem[{Schöier {et~al.}(2005)Schöier, van~der Tak, van Dishoeck, \&
  Black}]{schoier_atomic_2005}
Schöier, F.~L., van~der Tak, F. F.~S., van Dishoeck, E.~F., \& Black, J.~H.
  2005, Astronomy and Astrophysics, 432, 369,
  \dodoi{10.1051/0004-6361:20041729}

\bibitem[{Siess {et~al.}(2022)Siess, Homan, Toupin, \& Price}]{siess_3d_2022}
Siess, L., Homan, W., Toupin, S., \& Price, D.~J. 2022, {3D} simulations of
  {AGB} stellar winds -- {I}. {Steady} winds and dust formation, Tech. rep.
\newblock \url{https://ui.adsabs.harvard.edu/abs/2022arXiv220813869S}

\bibitem[{Stuart(2010)}]{stuart_inverse_2010}
Stuart, A.~M. 2010, Acta Numerica, 19, 451, \dodoi{10.1017/S0962492910000061}

\bibitem[{Tazzari {et~al.}(2018)Tazzari, Beaujean, \&
  Testi}]{tazzari_galario_2018}
Tazzari, M., Beaujean, F., \& Testi, L. 2018, Monthly Notices of the Royal
  Astronomical Society, 476, 4527, \dodoi{10.1093/mnras/sty409}

\bibitem[{{The Astropy Collaboration} {et~al.}(2013){The Astropy
  Collaboration}, Robitaille, Tollerud, Greenfield, Droettboom, Bray, Aldcroft,
  Davis, Ginsburg, Price-Whelan, Kerzendorf, Conley, Crighton, Barbary, Muna,
  Ferguson, Grollier, Parikh, Nair, Günther, Deil, Woillez, Conseil, Kramer,
  Turner, Singer, Fox, Weaver, Zabalza, Edwards, Bostroem, Burke, Casey,
  Crawford, Dencheva, Ely, Jenness, Labrie, Lim, Pierfederici, Pontzen, Ptak,
  Refsdal, Servillat, \& Streicher}]{the_astropy_collaboration_astropy_2013}
{The Astropy Collaboration}, Robitaille, T.~P., Tollerud, E.~J., {et~al.} 2013,
  Astronomy \& Astrophysics, 558, A33, \dodoi{10.1051/0004-6361/201322068}

\bibitem[{{The Astropy Collaboration} {et~al.}(2018){The Astropy
  Collaboration}, Price-Whelan, Sipőcz, Günther, Lim, Crawford, Conseil,
  Shupe, Craig, Dencheva, Ginsburg, VanderPlas, Bradley, Pérez-Suárez,
  Val-Borro, Contributors), Aldcroft, Cruz, Robitaille, Tollerud, Committee),
  Ardelean, Babej, Bach, Bachetti, Bakanov, Bamford, Barentsen, Barmby,
  Baumbach, Berry, Biscani, Boquien, Bostroem, Bouma, Brammer, Bray,
  Breytenbach, Buddelmeijer, Burke, Calderone, Rodríguez, Cara, Cardoso,
  Cheedella, Copin, Corrales, Crichton, D’Avella, Deil, Depagne, Dietrich,
  Donath, Droettboom, Earl, Erben, Fabbro, Ferreira, Finethy, Fox, Garrison,
  Gibbons, Goldstein, Gommers, Greco, Greenfield, Groener, Grollier, Hagen,
  Hirst, Homeier, Horton, Hosseinzadeh, Hu, Hunkeler, Ivezić, Jain, Jenness,
  Kanarek, Kendrew, Kern, Kerzendorf, Khvalko, King, Kirkby, Kulkarni, Kumar,
  Lee, Lenz, Littlefair, Ma, Macleod, Mastropietro, McCully, Montagnac, Morris,
  Mueller, Mumford, Muna, Murphy, Nelson, Nguyen, Ninan, Nöthe, Ogaz, Oh,
  Parejko, Parley, Pascual, Patil, Patil, Plunkett, Prochaska, Rastogi, Janga,
  Sabater, Sakurikar, Seifert, Sherbert, Sherwood-Taylor, Shih, Sick, Silbiger,
  Singanamalla, Singer, Sladen, Sooley, Sornarajah, Streicher, Teuben, Thomas,
  Tremblay, Turner, Terrón, Kerkwijk, Vega, Watkins, Weaver, Whitmore,
  Woillez, Zabalza, \& Contributors)}]{the_astropy_collaboration_astropy_2018}
{The Astropy Collaboration}, Price-Whelan, A.~M., Sipőcz, B.~M., {et~al.}
  2018, The Astronomical Journal, 156, 123, \dodoi{10.3847/1538-3881/aabc4f}

\bibitem[{{The Astropy Collaboration} {et~al.}(2022){The Astropy
  Collaboration}, Price-Whelan, Lim, Earl, Starkman, Bradley, Shupe, Patil,
  Corrales, Brasseur, Nöthe, Donath, Tollerud, Morris, Ginsburg, Vaher,
  Weaver, Tocknell, Jamieson, Kerkwijk, Robitaille, Merry, Bachetti, Günther,
  Authors, Aldcroft, Alvarado-Montes, Archibald, Bódi, Bapat, Barentsen,
  Bazán, Biswas, Boquien, Burke, Cara, Cara, Conroy, Conseil, Craig, Cross,
  Cruz, D’Eugenio, Dencheva, Devillepoix, Dietrich, Eigenbrot, Erben,
  Ferreira, Foreman-Mackey, Fox, Freij, Garg, Geda, Glattly, Gondhalekar,
  Gordon, Grant, Greenfield, Groener, Guest, Gurovich, Handberg, Hart,
  Hatfield-Dodds, Homeier, Hosseinzadeh, Jenness, Jones, Joseph, Kalmbach,
  Karamehmetoglu, Kałuszyński, Kelley, Kern, Kerzendorf, Koch, Kulumani, Lee,
  Ly, Ma, MacBride, Maljaars, Muna, Murphy, Norman, O’Steen, Oman, Pacifici,
  Pascual, Pascual-Granado, Patil, Perren, Pickering, Rastogi, Roulston, Ryan,
  Rykoff, Sabater, Sakurikar, Salgado, Sanghi, Saunders, Savchenko, Schwardt,
  Seifert-Eckert, Shih, Jain, Shukla, Sick, Simpson, Singanamalla, Singer,
  Singhal, Sinha, Sipőcz, Spitler, Stansby, Streicher, Šumak, Swinbank,
  Taranu, Tewary, Tremblay, Val-Borro, Kooten, Vasović, Verma, Cardoso,
  Williams, Wilson, Winkel, Wood-Vasey, Xue, Yoachim, Zhang, Zonca, \&
  Contributors}]{the_astropy_collaboration_astropy_2022}
{The Astropy Collaboration}, Price-Whelan, A.~M., Lim, P.~L., {et~al.} 2022,
  The Astrophysical Journal, 935, 167, \dodoi{10.3847/1538-4357/ac7c74}

\bibitem[{{The Casa Team} {et~al.}(2022){The Casa Team}, Bean, Bhatnagar,
  Castro, Meyer, Emonts, Garcia, Garwood, Golap, Villalba, Harris, Hayashi,
  Hoskins, Hsieh, Jagannathan, Kawasaki, Keimpema, Kettenis, Lopez, Marvil,
  Masters, McNichols, Mehringer, Miel, Moellenbrock, Montesino, Nakazato, Ott,
  Petry, Pokorny, Raba, Rau, Schiebel, Schweighart, Sekhar, Shimada, Small,
  Steeb, Sugimoto, Suoranta, Tsutsumi, Bemmel, Verkouter, Wells, Xiong,
  Szomoru, Griffith, Glendenning, \& Kern}]{the_casa_team_casa_2022}
{The Casa Team}, Bean, B., Bhatnagar, S., {et~al.} 2022, Publications of the
  Astronomical Society of the Pacific, 134, 114501,
  \dodoi{10.1088/1538-3873/ac9642}

\bibitem[{Thompson(1999)}]{thompson_fundamentals_1999}
Thompson, A.~R. 1999, 180, 11.
\newblock \url{https://ui.adsabs.harvard.edu/abs/1999ASPC..180...11T}

\bibitem[{Vicente~Arévalo {et~al.}(2022)Vicente~Arévalo, Asensio~Ramos, \&
  Esteban~Pozuelo}]{vicente_arevalo_accelerating_2022}
Vicente~Arévalo, A., Asensio~Ramos, A., \& Esteban~Pozuelo, S. 2022, The
  Astrophysical Journal, 928, 101, \dodoi{10.3847/1538-4357/ac53b3}

\bibitem[{Von~Zeipel(1908)}]{von_zeipel_catalogue_1908}
Von~Zeipel, H. 1908, Annales de l'Observatoire de Paris, 25, F.1.
\newblock \url{https://ui.adsabs.harvard.edu/abs/1908AnPar..25F...1V}

\bibitem[{Öberg {et~al.}(2021)Öberg, Guzmán, Walsh, Aikawa, Bergin, Law,
  Loomis, Alarcón, Andrews, Bae, Bergner, Boehler, Booth, Bosman, Calahan,
  Cataldi, Cleeves, Czekala, Furuya, Huang, Ilee, Kurtovic, Le~Gal, Liu, Long,
  Ménard, Nomura, Pérez, Qi, Schwarz, Sierra, Teague, Tsukagoshi, Yamato,
  van't Hoff, Waggoner, Wilner, \& Zhang}]{oberg_molecules_2021}
Öberg, K.~I., Guzmán, V.~V., Walsh, C., {et~al.} 2021, The Astrophysical
  Journal Supplement Series, 257, 1, \dodoi{10.3847/1538-4365/ac1432}

\bibitem[{Štěpán {et~al.}(2022)Štěpán, Alemán, \&
  Bueno}]{stepan_novel_2022}
Štěpán, J., Alemán, T. d.~P., \& Bueno, J.~T. 2022, Astronomy \&
  Astrophysics, 659, A137, \dodoi{10.1051/0004-6361/202142079}

\end{thebibliography}
\bibliographystyle{aasjournal}



\end{document}